\documentclass[12pt,notoc]{JHEP3}
\usepackage{amsmath,amssymb,euscript,array,mathrsfs}
\usepackage{epsfig}
\newcommand{\beq}{\begin{equation}}
\newcommand{\eeq}{\end{equation}}
\newcommand{\bea}{\begin{eqnarray}}
\newcommand{\eea}{\end{eqnarray}}
\newcommand{\beqn}{\begin{eqnarray}}
\newcommand{\eeqn}{\end{eqnarray}}
\newcommand{\beas}{\begin{eqnarray*}}
\newcommand{\eeas}{\end{eqnarray*}}

\newcommand{\bquo}{\begin{quote}}
\newcommand{\enqu}{\end{quote}}

\newcommand{\startappendix}{
\setcounter{section}{0}
\renewcommand{\thesection}{\Alph{section}}}
\newcommand{\Appendix}[1]{
\refstepcounter{section}
\begin{flushleft}
{\large\bf Appendix \thesection: #1}
\end{flushleft}}
\def\Tr{ \hbox{\rm Tr}}

\def\stroke{\vrule height8pt width0.4pt depth-0.1pt}
\def\topfleck{\vrule height8pt width0.5pt depth-5.9pt}
\def\botfleck{\vrule height2pt width0.5pt depth0.1pt}
\def\Zmath{\vcenter{\hbox{\numbers\rlap{\rlap{Z}\kern 0.8pt\topfleck}\kern
2.2pt\rlap Z\kern 6pt\botfleck\kern 1pt}}}
\def\Qmath{\vcenter{\hbox{\upright\rlap{\rlap{Q}\kern
3.8pt\stroke}\phantom{Q}}}}
\def\Nmath{\vcenter{\hbox{\upright\rlap{I}\kern 1.7pt N}}}
\def\Cmath{\vcenter{\hbox{\upright\rlap{\rlap{C}\kern
3.8pt\stroke}\phantom{C}}}}
\def\Rmath{\vcenter{\hbox{\upright\rlap{I}\kern 1.7pt R}}}
\def\Z{\ifmmode\Zmath\else$\Zmath$\fi}
\def\Q{\ifmmode\Qmath\else$\Qmath$\fi}
\def\N{\ifmmode\Nmath\else$\Nmath$\fi}
\def\C{\ifmmode\Cmath\else$\Cmath$\fi}
\def\R{\ifmmode\Rmath\else$\Rmath$\fi}
\def\Tr{{\rm Tr}}

\def\2{{1\over 2}}

\def\4N{${\cal N}=4$}

\def\beq{\begin{equation}}
\def\eeq{\end{equation}}
\def\ba{\beq\new\begin{array}{c}}
\def\ea{\end{array}\eeq}

\title{Non-Abelian $k$-Vortex Dynamics in ${\cal N}=1^*$ theory and its 
Gravity Dual}
\author{ Roberto~Auzzi  and  S.~Prem~Kumar\\
{\it 
Department of Physics, Swansea University, \\ Singleton Park,Swansea\\SA2 8PP, U.K.
}\\
E-mail: \email{r.auzzi@swansea.ac.uk}, \email{s.p.kumar@swansea.ac.uk}
} 
\abstract{We study magnetic flux tubes in the Higgs vacuum of
 the ${\cal N}=1^*$ mass deformation of $SU(N_c)$, ${\cal N}=4$ SYM and its 
large $N_c$ string dual, the Polchinski-Strassler geometry. 
Choosing equal masses for the three adjoint
chiral multiplets, for all $N_c$ we identify a ``colour-flavour
locked''  symmetry, $SO(3)_{C+F}$ which leaves the Higgs vacuum
invariant. At weak coupling, we find explicit non-Abelian
$k$-vortex solutions carrying a ${\mathbb Z}_{N_c}$-valued magnetic
flux, with winding, $ 0 < k  < N_c$. These $k$-strings
spontaneously break $SO(3)_{C+F}$ to $U(1)_{C+F}$ resulting 
in an $S^2$ moduli space of solutions. The world-sheet
sigma model is a nonsupersymmetric 
${\mathbb{CP}}^1$ model with a 
theta angle $\theta_{1+1} = k(N_c-k) \theta_{3+1}$ where $\theta_{3+1}$
is the Yang-Mills vacuum angle. We find numerically 
that $k$-vortex tensions follow the Casimir scaling law
$T_k \propto k (N_c-k)$ for large $N_c$. 
In the large $N_c$ IIB string
dual, the $SO(3)_{C+F}$ symmetry is manifest in the geometry interpolating
between $AdS_5\times S^5$ and the interior metric due to a single
D5-brane carrying D3-brane charge. We identify candidate $k$-vortices as
expanded probe D3-branes formed from a collection of $k$ D-strings. 
The resulting $k$-vortex tension exhibits precise Casimir scaling, and 
the effective world-sheet theta angle matches the
semiclassical result. S-duality maps the Higgs to the confining
phase so that confining string tensions at strong 't Hooft  coupling
also exhibit Casimir scaling in ${\cal N}=1^*$ theory in the large $N_c$ limit.
}
\begin{document}

\section{Introduction}

Supersymmetric gauge theories, following the works of \cite{sw}, have 
provided a large class of examples where condensation of monopoles
is the mechanism for confinement of electric charges.
 Softly broken ${\cal N}=2$ supersymmetric gauge
theories confine via a magnetic version of the Abelian Higgs
mechanism. In these theories the confined, heavy, coloured sources are
held together by Abelian strings (Abrikosov-Nielsen-Olesen solitons
\cite{ano,nielsen}). 
In contrast, in pure Yang-Mills theory with $SU(N_c)$ gauge
group for instance, heavy external charges are expected to be confined
by chromoelectric flux tubes which annihilate in groups of $N_c$,
which we refer to as $\mathbb{Z}_{N_c}$-strings.
One example where the dynamics of $\mathbb{Z}_{N_c}$ strings may be
accessed at weak coupling, is presented by the so-called ${\cal N}=1^*$ theory 
\cite{Vafa:1994tf,Donagi:1995cf,Dorey:1999sj,Dorey:2000fc, ps, strassler} 
which is a mass deformation of ${\cal N}=4$ theory preserving ${\cal N}=1$
supersymmetry (SUSY). What makes this theory particularly interesting
is that it also has a known large $N_c$ string dual \cite{ps} which brings with
it the possibility of exploring flux tube dynamics in the large $N_c$ limit.

The ${\cal N}=1^*$ theory has extremely rich infrared dynamics and
beautiful phase structure, made possible in part by the Olive-Montonen
electric-magnetic duality \cite{Montonen:1977sn}
(enlarged to $SL(2,{\mathbb Z})$) which it
inherits from its parent ${\cal N}=4$ theory. 
For example, the theory with $SU(N_c)$ gauge group has a large number
of vacua with a mass gap, each of which is realized in a distinct
phase. The action of $SL(2,{\mathbb Z})$ exchanges and permutes these
vacua. The vacuum in the Higgs phase, where the gauge group is broken
completely, is mapped by S-duality (inversion of the gauge coupling)
onto the confining vacuum where the gauge group is classically
unbroken and the theory confines in the infrared (IR).
The $\mathbb{Z}_{N_c}$ chromoelectric flux tubes in the confined phase 
at strong gauge coupling, get directly mapped to  magnetic
flux tubes in the Higgs vacuum at weak coupling. At weak coupling, the
Higgs vacuum is semiclassical and hence the physics of the associated 
flux tubes is accessible. The study of these for general
$N_c$, and particularly their 
large $N_c$ gravity duals , will be the subject of this paper.

In recent years, certain special flux tubes at weak coupling have been
encountered in gauge theories (with and without SUSY)
with $U(N_c)$ gauge
group and $N_f$ flavours in the fundamental representation
\cite{HT,ABEKY} and extensively studied therein
\cite{dt,Shifman:2004dr,HT2}. The crucial feature of all these strings
at weak coupling  is the presence of orientational moduli associated with 
rotations within a colour-flavour locked symmetry. We will refer to
these as ``non-Abelian'' flux tubes.
The interested reader can find reviews in \cite{SYrev,Trev1, Trev2,EINOKrev}.
 In the context of ${\cal N}=1^*$
theory with $SU(2)$ gauge group, non-Abelian vortices in the Higgs
vacuum were constructed and studied first in \cite{mamayung}.

The basic example of the non-Abelian strings is in the context of 
$\mathcal{N}=2$, $U(N_c)$ gauge theory with $N_f$
flavours and $N_f=N_c=N$ and a Fayet-Iliopolous term.
In this case there is an $SU(N)_{C+F}$ symmetry which is left unbroken by the vacuum;
the vortex soliton breaks this symmetry to $(SU(N-1) \times U(1))_{C+F}$.
The vortex internal space is then parameterized by
\beq
\mathcal{M}^{{\cal N} =2}= \frac{SU(N)_{C+F}}{(SU(N-1) \times
  U(1))_{C+F}} =\mathbb{CP}^{N-1}\, .\nonumber 
\eeq

In this paper we will study a similar ``colour-flavour locked''
symmetry that appears in the Higgs vacuum of $\mathcal{N}=1^*$ theory with
$SU(N_c)$ gauge group. When the masses of the three adjoint 
${\cal N}=1$ chiral multiplets in the theory are chosen to be equal,
an $SO(3)$ subgroup of the
original global $SO(6)_R$ symmetry of ${\cal N}=4$ theory is left
unbroken. The VEVs of the scalar fields in this phase are 
proportional to $N_c$ dimensional $SU(2)$ generators. This fact allows
to find a specific combination of the global $SO(3)$ and 
colour generators, that are left unbroken by the VEVs of
the adjoint scalars. We denote this 
combined colour-flavour symmetry as $SO(3)_{C+F}$.

Since all fields in the theory are in the adjoint representation of
the gauge group, the topologically
stable flux tubes are classified by a 
$\mathbb{Z}_{N_c}$  quantum number $k=1,2,\, .. \, N_c-1$
\footnote{Flux tubes at weak coupling with $\mathbb{Z}_{N_c}$ quantum numbers 
were also studied in numerous papers (see \cite{misc}, \cite{kneipp}
for an incomplete list).}. 
In the first part of this paper we find a general ansatz for the
$k$-vortex solution, generalizing 
the $N_c=2$, $k=1$ case studied in \cite{mamayung}.
The ansatz is given in explicit form for $N_c=3,4$ and a natural algorithm
for higher rank gauge groups presents itself. Since the equations of
motion are not analytically tractable, a numerical solution of the
vortex profile functions is necessary.  We were able to perform  the numerical
computations for $k$-vortices in theories with $2 \leq N_c \leq
6$. 

The $k$-vortex solution breaks the $SO(3)_{C+F}$ symmetry to $U(1)_{C+F}$,
so that the vortex internal moduli space (for every $k$) is parameterized by
\beq
\mathcal{M}^{{\cal N}=1^*}=\frac{SO(3)_{C+F}}{U(1)_{C+F}}
=\mathbb{CP}^1\, 
.\nonumber
\eeq
Acting on a given $k$-string solution with the broken symmetry
generators rotates the orientation of the non-Abelian 
magnetic flux within the colour space. A crucial difference between
the vortices in ${\cal N}=1^*$ theory and those in theories with ${\cal
N}=2$ SUSY, is that the latter are BPS solutions. 
With $SO(3)$ symmetric masses, the ${\cal N}=1^*$ vortices
are far from BPS
\footnote{ There is a limiting regime of mass parameters
(two masses equal, and the third being relatively small) where the
${\cal N}=1^*$ theory can be viewed as softly broken ${\cal N}=2^*$
theory, but we will not be particularly interested in this
limit. Abelian vortices in softly broken ${\cal N}=2^*$ theory are
BPS.} and have no fermionic ``super-orientational'' zero modes.

The low-energy effective theory for the fluctuations of the light
modes on the $k$-string is determined by performing an adiabatic, world-sheet 
dependent  colour-flavour locked rotation. This excites the internal, 
orientational zero mode   
degrees of freedom localised
on the vortex. The resulting action is that of a sigma model with
$S^2$ as target space and the following Lagrangian, 
\beq  S_{1+1} = \int dz \, dt \,  
\left(  B_{N_c,k} (\partial_s \vec{n})^2 
 -\frac{\theta^{N_c,k}_{1+1}}{8 \pi}
 \epsilon^{s r} \epsilon_{a b c} n^a \partial_s n^b \partial_r n^c
    \right) \, , \label{effeti} \eeq
where $\vec n$ is a position vector on the unit sphere. This is an
effective theory with a UV cutoff determined by the vortex thickness. 
Importantly, the effective theory is an asymptotically free quantum
theory and its IR dynamics depends strongly on the vacuum theta angle
\cite{Shifman:2004dr, witten, zam1, zam2, zam3}. Therfore, while the
four dimensional gauge theory is semiclassical, the vortex theory is
highly quantum and becomes strongly interacting.
The classical
value of the sigma model coupling  $B_{N_c,k}$ can be determined in
terms of the Yang-Mills coupling $g^2_{\rm YM}$,
 for all $k$ and $N_c$, by a non-trivial
 numerical calculation involving the vortex profile functions. The
 classical sigma model coupling constant turns out to be weak for weak
 gauge coupling.
On the other hand, the effective 2-dimensional $\theta$ angle can be
computed analytically in general, to yield a simple, but very
interesting result,
\beq 
\theta^{N_c,k}_{1+1}=k (N_c-k) \theta_{3+1} \, \label{the},
\eeq
where $\theta_{3+1}$ is the vacuum angle of the four dimensional gauge
theory. This relation is significant for two reasons. First, it
satisfies the basic requirement that the Higgs vacuum should be
invariant under shifts of $\theta_{3+1}$ by multiples of
$2\pi$. Second, whenever $\theta_{1+1}=\pi$, the world-sheet theory is
integrable and flows to a conformal fixed point with massless $SO(3)$
doublets as the only excitations. For all other values of
$\theta_{1+1}$ the two dimensional theory develops a mass gap and its
only excitations are triplets of $SO(3)$ which may be viewed as
confined meson-like states made up of doublets. This in turn implies 
that there exist various special values for $\theta_{3+1}$, determined
by (\ref{the}) for every $k$, at which different $k$-vortex theories
flow to an interacting conformal fixed point with central charge $c=1$. 

Since we find the explicit $k$-vortex solutions, albeit numerically,
we are in a position to ask how their tensions scale with $N_c$. This
is a question that has attracted considerable interest in recent
years, from various perspectives \cite{strassler,biaggio,hk,ds,hsz}
for gauge theories with a ${\mathbb Z}_{N_c}$ symmetry. 
We perform a numerical analysis of the semiclassical $k$-string
tensions and their 
ratios for $N_c=4,5,6$. We find that as $N_c$ is increased, the
results are extremely well approximated by a Casimir scaling
law with an accuracy better than $0.1 \%$. Although we do not
yet have an understanding of the physics behind this result, we are
able to confirm that Casimir scaling of the tensions becomes precise in
the large $N_c$ gravity dual. At this point it is worth emphasizing that
S-duality maps these Higgs phase results at $g^2_{YM}\ll 1$ to
the confining vacuum at $g^2_{YM}\gg1$.

The second part of our paper is devoted to a study  of $k$-strings in
the Higgs vacuum in the large $N_c$, Type IIB string dual obtained by 
Polchinski and Strassler \cite{ps}. The supergravity background which is dual to the
Higgs vacuum, becomes applicable when $N_c\rightarrow\infty$ and $N_c/g^2_{\rm YM}\gg
1$. Since this also includes the regime of weak gauge coupling, we
cannot expect supergravity to be valid in the entire geometry as
a weakly coupled regime would correspond to large curvatures in the
string dual. This also occurs in the Higgs vacuum, where the dual
background interpolates between $AdS_5\times S^5$ asymptotics and a
deep interior portion generated by a D5-brane wrapped on a flux
supported two-sphere. The D5-brane which carries $N_c$ units of D3-brane charge,
makes an appearance due to the Myers effect \cite{myers}
resulting from the
${\cal N}=1^*$ deformation. In the crossover region, near the D5-brane, the geometry
becomes strongly curved and we expect large string corrections. 
Despite this we can certainly look for candidate probe brane configurations
that are expected to be dual to the $k$-vortices of the Higgs vacuum
in the large $N_c$ limit. By
S-duality, the picture in the Higgs vacuum is exchanged with the
confining vacuum at strong 't Hooft coupling: $N_c/g^2_{\rm YM}$ $\to$
$g^2_{\rm YM} N_c\gg 1$ which is the usual condition for the validity
of supergravity.

With all the above caveats in mind, we look for our candidate probe
branes in the dual geometry. The $SO(3)_{C+F}$ is obvious in the
geometry as the sphere wrapped by the D5-brane has an $SO(3)$ isometry 
in the limit of equal masses for the adjoint chiral multiplets.
The $k=1$ vortex is naturally a probe D1-brane 
in the Higgs vacuum. In the brane picture, the D1-brane binds to
the D5-brane which has a world-volume $B$-field endowing the 5-brane
with D3 charge. This bound state corresponds to a magnetic flux
tube. In the gravity picture, the probe D1-brane sits at a radial
position near the D5-brane. Despite the possibility of stringy
corrections to the background, we use the probe 
Dirac-Born-Infeld  action and the Chern-Simons terms
to obtain the effective Lagragian in Eq. (\ref{effeti}), with 
\beq B_{N_c,1}=\frac{\pi N_c}{ g^2_{\rm YM}}, \qquad \theta^{N_c,1}_{1+1} =  N_c \theta_{3+1} \, .\eeq 
The value of $ \theta^{N_c,1}_{1+1}$ which is found in the dual
is consistent with Eq.~(\ref{the}) for large $N_c$. The tension of
this configuration can also be computed (as originally done in
\cite{ps}) and yields $T_{k=1}= 2\pi m^2 N_c/g^2_{\rm YM}$.

In order to model the $k$-string with $k \sim {\cal O}(N_c)$, 
motivated by the Myers dielectric effect on a collection of $k$
D-strings in the Higgs vacuum,  we use a D3-brane with 
the topology of $\mathbb{R}^{1,1} \times S^2$, with 
$k$ units of flux in the compact directions. Crucially, the primary
contribution to the tension of this D3-brane is a disc stretching
inside the D5-sphere, a picture that we find to be consistent with the
baryon vertex in ${\cal N}=1^*$ theory. 
From the D3-brane picture we find that the vortex tension follows 
the Casimir scaling law 
\beq T_k=2 \pi \frac{ m^2}{ g_{\rm YM}^2} k (N_c-k) \, .\eeq
reproducing precisely the semiclassical field theory result which was
determined numerically. Most remarkably the Chern-Simons terms of the probe
brane also compute the theta angle on the $k$-vortex worldsheet,
exactly matching Eq.~(\ref{the}), the weak coupling gauge theory result.

The agreement between the gravity dual and semiclassical gauge theory
physics is surprising and clearly needs an explanation. An important
aspect of the probe brane results from the string dual is that the
physical quantities that agree with the gauge theory - the $k$-string
tension and the worldsheet theta angle - do not appear to 
receive significant contributions from the strongly curved parts of
the geometry. The D3-brane $k$-string tension arises mainly from a
disc-like portion that effectively sees a flat geometry inside
the D5-sphere, while the theta angle originates in the Chern-Simons
term which is insensitive to the metric. Therefore, we believe that
the above picture and results are robust.

The paper is organized as follows. In Sect. \ref{setting} we review
certain aspects of the ${\cal N}=1^*$ field theory.  
In Sect.~\ref{solitons} we present solitonic solutions
for the $k$-strings and present the results of numerical analysis. 
In Sect.~\ref{effective} we derive the vortex world-sheet effective action
from a direct calculation. Sect.~\ref{sugra} deals 
with the probe brane calculation in the Polchinski-Strassler
background for the Higgs vacuum. 
Sect.~\ref{summary} briefly summarizes our conclusions.
Some details of the interior geometry
in the Polchinski-Strassler background are presented in an Appendix.

\section{The Field Theory setting} 
\label{setting}
In this section we cover some of the basic facts regarding the ${\cal
  N}=1^*$ field theory with $SU(N_c)$ gauge group.  We pay particular attention
  to the theory in the Higgs phase, and for a specific choice of the mass
deformation  parameters. The physics in this vacuum is related directly via
  S-duality to the confining phase of the theory.

\subsection{The ${\cal N}=1^*$ deformation of ${\cal N}=4$ SYM}

We begin by reviewing the field content and the microscopic Lagrangian of the 
${\cal N}=1^*$ theory. In the language of $\mathcal{N}=1$ supersymmetry,
the $\mathcal{N}=1^*$ theory contains  
an ${\cal N}=1$ vector multiplet $W_\alpha$ 
and three chiral multiplets $(\Phi_1, \Phi_2, \Phi_3)$, transforming in the
adjoint representation of the gauge group which we take to be
$SU(N_c)$. The theory is obtained by a relevant, mass deformation of
${\cal N}=4$ supersymmetric Yang-Mills theory.
The superpotential of $\mathcal{N}=4$ SYM reads,
\beq \mathcal{W}={1\over g_{\rm YM}^2} \;\Tr ([\Phi_1,\Phi_2] \Phi_3). \eeq
The superpotential can be deformed by adding ${\cal N}=1$ SUSY preserving 
mass terms for the adjoint matter fields,
\beq 
\Delta \mathcal{W}= \frac{1}{ g_{\rm YM}^2} \sum_{i=1}^3  \frac{1}{2}
{m_i} \Tr(\Phi_i^2).
\eeq 
This is a relevant deformation and the resulting theory exhibits
nontrivial dynamics in the infrared, resulting in a rich phase
structure. In the UV however, the theory flows to ${\cal N}=4$ SYM,
with the gauge coupling remaining a freely adjustable parameter. Thus the
${\cal N}=1^*$ theory has, in addition to three complex mass
parameters, a dimensionless, tunable complexified gauge coupling
\beq
\tau= {4\pi i\over g^2_{\rm YM}} + {\theta_{3+1}\over 2\pi}.
\eeq

In Euclidean space, the bosonic part of the action is,
\beq 
S_{\rm b}^{\rm E}=\int d^4 x \; \left[\frac{1}{g_{\rm YM}^2}\;  \left(
  \frac{1}{4} 
F^a_{\mu \nu} F^{a\mu\nu}+
\sum_{j=1}^3\Tr |D_\mu \Phi_j|^2 + V_D + V_F
\right)+  \frac{i \, \theta_{3+1}}{32 \pi^2} F^a_{\mu \nu}
\tilde{F}^a_{\mu \nu}\right] \, ,
\eeq
where we have used the same symbol $\Phi_j$,
for the chiral superfields
as for their lowest (scalar) components. We define the 
$SU(N_c)$ generators $T^a$,
$(a=1,2,\ldots N^2_c-1)$, (with $F_{\mu \nu}=T_a F^{a}_{\mu \nu}$), 
with the usual normalization $\Tr (T_a T_b)=\frac{1}{2} \delta_{ab}$,
while the gauge covariant derivative is
\beq D_\mu \Phi_k=\partial_\mu \Phi_k - i [A_\mu,\Phi_k].\eeq 
The scalar potential is the sum of $V_F$ and $V_D$, the F and D-term
contributions respectively:
\beq
V_F = \Tr \left( w_1.w_1^\dagger + w_2.w_2^\dagger
+w_3.w_3^\dagger \right)\,, \qquad w_i= \epsilon_{ijk}\Phi_j\Phi_k+ m_i\Phi_i\,,
\eeq
and
\beq V_D=-\frac{1}{4}\Tr \left( [\Phi_1^\dagger,\Phi_1] + [\Phi_2^\dagger,\Phi_2] +
[\Phi_3^\dagger,\Phi_3]   \right)^2\,.
\eeq
%Note that in the $\mathcal{N}=4$ limit, for $m_k=0$,
%the potential can be written as:
%\beq V=V_D+V_F= \sum_{1\leq i<j \leq 6} \Tr [\phi_i,\phi_k]^2,\eeq
%where  $\phi_i$ are the real and imaginary part of the complex
%fields $\Phi_k$.   

In this paper we will be mainly interested in the case where the
masses of the three adjoint chiral multiplets are equal: 
\beq
m_1=m_2=m_3=m.
\eeq
With this choice, the superpotential term $\Delta \mathcal{W}$ breaks
the $SO(6)_R$ global symmetry of the $\mathcal{N}=4$ theory
to an $SO(3)$ subgroup under which the complex chiral multiplets 
$(\Phi_1,\Phi_2,\Phi_3)$, transform as a triplet. In the ${\cal N}=1^*$ theory, this
 $SO(3)$ acts as an ordinary global symmetry, and not as an
 R-symmetry.

\subsection{Higgs and Confining Vacua}
The mass deformation above results in a large set of vacuum
configurations determined by the F-flatness conditions (modulo complex
gauge transformations),
\beq
\Phi_i= -{1\over m}\,\epsilon_{ijk}\,\Phi_j\Phi_k.
\eeq
As is well-known \cite{Donagi:1995cf,Dorey:1999sj}, 
the solutions to these equations may be classified
in terms of all $N_c$-dimensional representations the $SU(2)$
algebra. Each such classical ground state then splits into a certain
number quantum vacua depending on the non-Abelian gauge symmetry
subgroup left unbroken by the classical solution. The quantum ground
states are in one to one correspondence with all possible phases of
$SU(N_c)$ gauge theory with adjoint matter, in four dimensions.

Of
particular interest are the Higgs and confining vacua which correspond
to the 
$N_c$ dimensional ireducible representation and the trivial
representation, respectively. 
The VEVs of the adjoint scalars in the Higgs vacuum are proportional to 
the generators of the irreducible $SU(2)$ representation with dimension $N_c$,
\beq \Phi_l = i \,m \,J_l\,,\qquad(l=1,2,3).\eeq
For generic $N_c$, the $SU(2)$ representation is labelled by 
$j=\frac{N_c-1}{2}$ with $J_3$ chosen to be the usual diagonal matrix
\beq J_3={\rm diag} (j, j-1, \ldots, -j)\,;\qquad j=\frac{N_c-1}{2}. \eeq
The only non-zero elements of the matrices $J_1$, $J_2$ are off-diagonal, 
given by 
\begin{eqnarray}
&&(J_1)_{a,\,a+1} = (J_1)_{a+1,\,a} = \frac{\sqrt{a(N_c-a)}}{2} \, , 
\qquad a=1,2,\ldots N_c-1
\\\nonumber
&&(J_2)_{a,\,a+1} = -(J_2)_{a+1,\,a} = - i \frac{\sqrt{a(N_c-a)}}{2} \, .
\end{eqnarray}
The usual relation between the generators of the $SU(2)$ algebra and
 the quadratic Casimir then follows,
\beq J_1^2+J_2^2+J_3^2=j(j+1) {\bf 1}
=\frac{(N^2_c-1)}{4}{\bf 1}\,.\eeq
This relation leads to a natural association
 of the Higgs vacuum of ${\cal N}=1^*$ theory with fuzzy sphere
 configurations of D3 branes     
in the string theory dual \cite{ps}.

The results we deduce below for magnetic flux tubes in the Higgs
vacuum, will have a direct bearing on the tension of the chromoelectric
flux tubes in the confining vacuum. This is because the
$SL(2,{\mathbb Z})$ electric-magnetic duality of the parent ${\cal
  N}=4$ theory permutes different IR phases of the ${\cal N}=1^*$
theory \cite{Vafa:1994tf, Donagi:1995cf, Dorey:1999sj}. 
In particular, the Higgs and confing vacua are exchanged
under S-duality: $\tau \rightarrow - 1/\tau$.
% Let us write down these matrices in some simple examples. For $N=2$
% the VEVS are proportional to the Pauli matrices:  
% \beq S_1= \frac{1}{2} \left(\begin{array}{ccc} 0  & 1 \\ 1 & 0 \\
% \end{array}\right), \,\,\,
% S_2= \frac{1}{2} \left(\begin{array}{ccc}
% 0  & -i \\i & 0  \\ \end{array}\right), \,\,\,
% S_3= \frac{1}{2} \left(\begin{array}{cc} 1  & 0 \\ 0 & -1  \\
% \end{array}\right) \eeq
% For $N=3$, the  VEVs  are proportional to:
% \beq  S_1= \frac{1}{\sqrt{2}} \left(\begin{array}{ccc} 0  & 1 & 0\\ 1 & 0 & 1 \\0 & 1 & 0 \\
% \end{array}\right), \,\,\, S_2=
% \frac{1}{\sqrt{2}} \left(\begin{array}{ccc} 0  & -i & 0\\ i & 0 & -i \\ 0 & i & 0 \\
% \end{array}\right), \,\,\,
% S_3=  \left(\begin{array}{ccc} 1  & 0 & 0\\ 0 & 0 & 0 \\ 0 & 0 & -1 \\
% \end{array}\right)\eeq
% For $N=4$: \beq 
% S_1= \frac{1}{2} \left(\begin{array}{cccc} 0  & \sqrt{3} & 0 & 0\\ \sqrt{3} & 0 & 2 & 0 \\ 0 & 2 & 0 & \sqrt{3} \\
% 0 & 0 & \sqrt{3} & 0\\ \end{array}\right), \,\,\,
% S_2= \frac{1}{2} \left(\begin{array}{cccc} 0  & - i \sqrt{3} & 0 & 0\\ i \sqrt{3} & 0 & - i 2 & 0 \\
% 0 & i 2 & 0 & -i \sqrt{3} \\ 0 & 0 & i \sqrt{3} & 0\\
% \end{array}\right), 
% \eeq \[ S_3= \frac{1}{2} \left(\begin{array}{cccc} 3  & 0 & 0 & 0\\ 0 & 1 & 0  & 0\\
% 0 & 0 & -1  & 0\\ 0 & 0 & 0  & -3\\ \end{array}\right).\]

\subsection{Colour-Flavour locking}
The VEVs of the adjoint scalars in the Higgs vacuum break 
the $SO(3)$ global  symmetry and the $SU(N_c)$ gauge symmetry.
However, it is always possible to find a combined global colour-flavour
rotation  which is unbroken \cite{mamayung}. This combined
$SO(3)_{C+F}$ global symmetry subgroup can be understood as follows. 
Any global $SO(3)$ rotation of the triplet $(\Phi_1,\Phi_2,\Phi_3)$
can be undone by a global colour transformation whose 
generators are chosen to be proportional to the
VEVs of the adjoint scalars i.e.,  the $N_c$ dimensional $SU(2)$ generators.
More explicitly, we first rotate the triplet $(\Phi_1,\Phi_2,\Phi_3)$
with the flavour matrix $U_F=\exp (T_j a_j)$,
where $T_j$ are the following $SO(3)$ generators,
\beq T_1=\left(\begin{array}{ccc}
0  & 0 & 0\\
0 & 0 & 1 \\
0 & -1 & 0 \\
\end{array}\right), \,\,\,
T_2=\left(\begin{array}{ccc}
0  & 0 & -1\\
0 & 0 & 0 \\
1 & 0 & 0 \\
\end{array}\right), \,\,\,
T_3=\left(\begin{array}{ccc}
0  & 1 & 0\\
-1 & 0 & 0 \\
0 & 0 & 0 \\
\end{array}\right). \eeq
This transformation acts in the flavour space as
\beq
\vec{\Phi} \rightarrow U_F\,\vec{\Phi}\,.
\label{flavour}
\eeq
Then let us introduce the global colour matrix
$W_C=\exp (i J_l a_l)$, acting as:
\beq
\Phi_i \rightarrow W_C\,\Phi_i \,W_C^\dagger,
\label{colour}
\eeq
where $J_l$ are the $N_c$ dimensional representations of $SU(2)$ generators.

A combination of  the above flavour and colour rotations
are unbroken by the scalar VEVs. The existence of this $SO(3)_{C+F}$
symmetry allows the determination of the worldsheet sigma model of
vortices (magnetic flux tubes) in the Higgs vacuum, as we see below.

\subsection{Higgs Vacuum Spectrum}

The perturbative spectrum in the Higgs vacuum for $N_c=2$  
is given by an $SO(3)_{C+F}$ triplet of massive
vector $\mathcal{N}=1$ multiplets with mass $\sqrt{2} m$,
one chiral multiplet with mass $m$ and 5 chiral multiplets
with mass $2 m$. 
For general $N_c$ the perturbative
spectrum has been computed in \cite{da}.
The result is the following:
 there are always 3 massive
vectors multiplets with mass $\sqrt{2} m$ and 
one chiral multiplet 
\footnote{In \cite{da}, the spectrum of the $U(N_c)$ gauge theory was
  determined, which differs slightly from the $SU(N_c)$ theory
  discussed here. In particular, for $U(N_c)$, there are three
  additional chiral multiplets with mass $m$ and one massless vector multiplet.
}
with mass $m$.
In addition, for every $k=2,\ldots,N_c-1$
there are $4 k$ massive chiral multiplets with mass
$k m $, $2k + 1$ massive vectors with mass 
$\sqrt{k(k+1)} m $ and lastly, a set of  $2N_c+1$ chiral multiplets with
mass $N_c m$. 
For every $N_c$ all these particles fit in representations
of $SO(3)_{C+F}$.

A beautiful feature of the Higgs vacuum of the ${\cal N}=1^*$ theory 
is that in the large $N_c$ limit it provides a deconstruction of
a six dimensional theory compactified on a sphere. In particular, as 
discussed in \cite{da}, the perturbative spectrum of the $U(N_c)$,
${\cal N}=1^*$ theory,
is identical to the spectrum of the Maldacena-Nu${\tilde {\rm n}}$ez twisted
compactification 
of the ${\cal N}=(1,1)$ 
six dimensional $U(1)$ gauge theory on a two-dimensional sphere. This
interpretation is a direct consequence of the association of the Higgs
vacuum with a fuzzy sphere configuration \cite{madore} as described above.

\section{The ${\mathbb Z}_{N_c}$ vortex as a soliton} \label{solitons}

\subsection{General discussion}
Since the ${\cal N}=1^*$ theory has only fields transforming in the
adjoint representation of the gauge group, the Lagrangian is invariant
under transformations in the center ${\mathbb Z}_{N_c}$ of
$SU(N_c)$. In the Higgs vacuum, magnetic charges valued in ${\mathbb
  Z}_{N_c}=\pi_1\left[SU(N_c)/{\mathbb Z}_{N_c}\right]$ are confined
by magnetic flux tubes, also carrying a ${\mathbb Z}_{N_c}$ charge.
Since the fluxes are defined modulo $N_c$, they annihilate in groups
of $N_c$.
At weak coupling $g_{\rm YM}\ll 1$, 
the physics in the Higgs vacuum is semiclassical and the magnetic flux
tubes should be understood as ordinary non-Abelian vortex string
solutions of the classical equations of motion.

In this section we introduce an ansatz for solitonic $k$-strings
in the Higgs vacuum of the $\mathcal{N}=1^*$ theory with gauge group
$SU(N_c)$ in the semiclassical limit $g_{\rm YM}\ll 1$.
We  will write the ansatz  explicitly for $N_c=2,3,4$.  The resulting
vortices carry magnetic flux $k=1,2,\dots,{N_c-1}$, defined modulo $N_c$.
Since they annihilate in groups of $N_c$, $N_c=4$ is the minimal
gauge group for which a non-trivial $k=2$ string appears. There is
also a $k=2$ vortex for $N_c=3$, but is essentially equivalent to the 
$k=1$ vortex under the transformation $k \rightarrow N_c-k$.  

The generalization of our ansatz to general $k$ and $N_c$ is
straightforward, but explicit calculations with these ans\"atze get
quite complicated. For generic $N_c$ and $k$,
in order to solve 
the equations of motion of the gauge theory, we need to introduce an ansatz
which depends on $3(N_c-1)$ independent profile functions for the vortex.
We have performed explicit numerical computations for $2 \leq N_c \leq 6 $ 
and generic $k$.

The classical equations of motion for the bosonic fields read
\beq \partial_\mu F^{\mu \nu} - i [A_\mu,F^{\mu \nu}]=
{i\over 2} \sum_{l=1}^3 \left([D^\nu \Phi_l,\Phi_l^\dagger] 
 + [D^\nu\Phi_l^\dagger, \Phi_l]\right)
\, , \label{eqm} \eeq
\beq D^\mu D_\mu \Phi_i=  (m \, w_i^\dagger 
 - \epsilon_{i j l} [w_j^\dagger,\Phi_l^\dagger]) +
{\partial V_D\over \partial \Phi_i}
\, . \label{eqm2}\eeq
Below we list explicit vortex solutions which satisfy 
\beq
\Phi_i=-\Phi_i^\dagger\,,\qquad i=1,2,3.
\eeq
so that the D-term contribution to the potential is identically zero
when evaluated on the solution, and the resulting equations of motion are
somewhat simpler.
For general $N_c$, there are $N_c-1$ distinct 
topological sectors labelled by an integer 
$k$ with $0<k<N_c$.

The vortex configurations have the adjoint scalars $\Phi_i$
approaching, at infinity,  a gauge transform of their VEVs in the
Higgs vacuum. In particular, certain matrix
elements of the adjoint scalars undergo a $2\pi$ phase rotation upon
winding once around the vortex. This phase rotation corresponds to a
gauge transformation (at infinity) 
which is single-valued in $SU(N_c)/{\mathbb Z}_{N_c}$.
In our ans\"atze below, the solutions with winding number $k=1$ will
have the scalars winding at infinity, effectively generated by
\beq Y_1= \frac{1}{N_c} \;{\rm Diag} (1, \cdots, 1, - (N_c-1) ),  \eeq
resulting in a chromomagnetic flux proportional to $Y_1$. Thus the
flux picks out a specific direction in colour-flavour space and the associated
string is truly non-Abelian.

In these solutions, $\Phi_3$ is chosen to have no azimuthal variation,
whilst both $\Phi_1$ and $\Phi_2$ have nontrivial angular dependence, away from
the vortex core. In
particular, as the azimuthal angle $\varphi$ varies from $0$ to $2\pi$,
the  components $(\Phi_{1,2})_{N_c-1,N_c}$ 
wind with a phase $e^{i \varphi}$,
while $(\Phi_{1,2})_{N_c,N_c-1}$ wind with the opposite 
phase $e^{-i \varphi}$. 

For generic winding $1<k<N_c$ the flux carried by the corresponding
$k$-vortex is
proportional to 
\beq Y_k= {\rm Diag} \left(
\underbrace{\frac{k}{N_c} , \cdots, \frac{k}{N_c}}_{N_c-k\,{\rm
    elements}} ,\;\; - 
\frac{N_c-k}{N_c}, \cdots, -\frac{N_c-k}{N_c} \right), \eeq
and away from the vortex core, the adjoint scalars behave as
\beq
\Phi_{1,2}(r, \varphi) = e^{i Y_k \varphi} 
\;\Phi_{1,2}(r,\varphi=0)\;e^{-i Y_k\varphi}
\eeq
while $\Phi_3$ has only a radial dependence.
Under the effect of this 
rotation, the components $(\Phi_{1,2})_{N_c-k,N_c-k+1}$ and
$(\Phi_{1,2})_{N_c-k+1,N_c-k}$ wind around the vortex with phase 
$e^{i\varphi}$ and $e^{-i\varphi}$ respectively.  
We believe that these are the solutions of lowest tension
in each topological sector $k$, as each field winds
at infinity exactly once. The solutions also display an obvious
vortex/anti-vortex  symmetry, which is evident under the replacement 
$k \rightarrow N_c-k$.

The explicit vortex solutions will break the $SO(3)_{C+F}$
global symmetry. However, they are invariant under the action of a
$U(1)$ subgroup corresponding to global rotations acting on
$(\Phi_1,\Phi_2)$. The action of the broken global symmetry
generators  then leads to a 
$ SU(2)/U(1)\simeq {\mathbb {CP}}^1$
moduli space of solutions for generic $(N_c,k)$.

\subsection{$N_c=2$}

For $SU(2)$ gauge group, the vortex solutions were first found in 
\cite{mamayung}. Here we rederive their result for completeness,
\begin{eqnarray}
&&\Phi_1= \frac{i m}{2} \psi_1(r)\left(\begin{array}{cc}
0  & e^{i \varphi} \\
 e^{-i \varphi}  & 0 \\
\end{array}\right), \,\qquad
 \Phi_2=
\frac{i m }{2}\psi_1(r) \left(\begin{array}{cc}
0  & -i e^{i \varphi} \\
i e^{-i \varphi} & 0  \\
\end{array}\right), \\\nonumber\\\nonumber
&&\Phi_3= \frac{i m }{2 } \kappa_1(r)\left(\begin{array}{cc}
1  & 0 \\
0 & -1 \\
\end{array}\right),
\end{eqnarray}
where $\psi_1$ and $\kappa_1$ are profile functions to be determined
by the equations of motion. Both approach unity as $r\to \infty$, in
order to match up with the Higgs VEVs. Near the origin $\psi_1$
vanishes so that the solution is smooth at $r=0$.
It is obvious that this configuration will be invariant under a
combination of an $SO(2)$ flavour rotation acting on the pair
$(\Phi_1,\Phi_2)$ and a global colour rotation generated by
$\sigma_3$. Hence the full $SO(3)_{C+F}$ is broken to $U(1)_{C+F}$.
  
The gauge field is solved by a typical vortex form
\beq 
A_x=\frac{-y}{r^2} (1-f(r)) Y, \qquad
A_y=\frac{x}{r^2} (1-f(r)) Y,\quad Y = {1\over 2}\sigma_3.
\eeq
This picks out a direction in the colour space and results in a
magnetic flux also proportional to $Y$,
\beq
F_{xy}=-{f'(r)\over r}\, Y.
\eeq
\begin{figure}[h]
\begin{center}
\epsfig{file=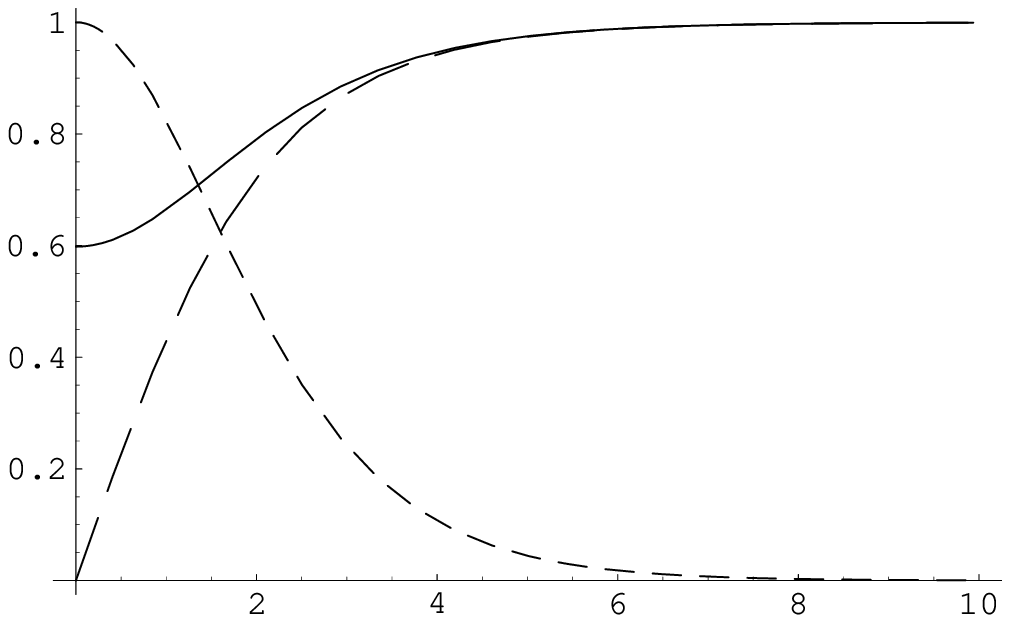, width =3.0in}
\end{center}
\noindent\small{\bf Figure 1:} The vortex profile functions  for $N_c=2$: $\kappa_{1}$ (solid),
$\psi_1$ (long dashes), $f$ (short dashes).
\label{nc2}
\end{figure}

The ansatz which is axisymmetric under rotations about the
$z$-axis, can be used to evaluate the action functional per unit
length. This yields the vortex tension functional 
\begin{eqnarray} 
&&T=\\\nonumber
&&2 \pi \int r \, dr \left( 
\frac{f'^2}{2 r^2}+ \frac{m^2 \kappa_1'^2}{2}+m^2 \psi_1'^2+
\frac{m^2 \psi_1^2 f^2}{ r^2}+ 
%\right.\\\nonumber
%&&\left. 
\frac{m^4}{2} ( (\kappa_1-\psi_1^2)^2 + 2 \psi_1^2 (\kappa_1-1)^2)
\right) .
\end{eqnarray}
It is easily checked that the equations of motion for the profile
functions
that follow from
varying this tension functional are the same as those following from
(\ref{eqm}) and (\ref{eqm2}). 
The profile functions are thus determined by solving, 
\begin{eqnarray} 
&&f''-\frac{f'}{r}-f \psi_1^2 2 m^2=0,\\
&&\psi_1'' + \frac{\psi_1'}{r}-\frac{\psi_1 f^2}{r^2}=
m^2 \psi_1 (\psi_1^2 +\kappa_1^2
-3 \kappa_1 +1) ,\\
&&\kappa_1''+\frac{\kappa_1'}{r}= m^2  (2 \psi_1^2 \kappa_1 -
3 \psi_1 ^2+\kappa_1).
\end{eqnarray}
The above equations can be solved numerically and the results are
plotted in Fig. 1. We learn from the solution that $\psi_1$ grows
from zero at the core of the vortex to unity at infinity. At the same
time, $\kappa_1$ remains non-zero at the string core whilst approaching
$1$ asymptotically. Hence, in the core $\Phi_3\neq0$,
$\Phi_1=\Phi_2=0$, so that there is a Coulomb-like
phase, shielded by a crossover region which eventually merges with the Higgs phase
vacuum at infinity. The profile function $f(r)$ shows that 
the magnetic field $\sim f'(r)/r$ is non-zero in the Coulomb-like
phase,  concentrated in a 
neighbourhood of the origin, while vanishing in the asymptotic 
Higgs vacuum.

\subsection{$N_c=3$}
Having understood the structure of the ${\mathbb Z}_2$ string for
$SU(2)$, we can now apply our general non-Abelian string ansatz
described in Section 3.1, to higher rank gauge groups. 
For $SU(3)$, and for $k=1$ our general ansatz takes the form,
\begin{eqnarray}
&&\Phi_1= \frac{i m}{\sqrt{2} } \left(\begin{array}{ccc}
0  & \psi_1  & 0\\
 \psi_1 & 0 & \psi_2 e^{i \varphi} \\
 0 & \psi_2 e^{-i \varphi}    & 0\\
\end{array}\right),  \,\qquad
\Phi_2= \frac{i m}{\sqrt{2} } \left(\begin{array}{ccc}
0  & -i\psi_1 & 0\\
 i\psi_1  & 0 & -i\psi_2 e^{i \varphi} \\
 0 & i\psi_2 e^{-i \varphi}    & 0\\
\end{array}\right), \\\nonumber\\\nonumber
&&\Phi_3= i m  \left(\begin{array}{ccc}
\kappa_1 -\kappa_2 /2 & 0  & 0\\
0 & \kappa_2   & 0\\
0 & 0   &  -\kappa_1 -\kappa_2/2  \\
\end{array}\right).
\end{eqnarray}
The forms of $\Phi_1$ and $\Phi_2$ are both motivated  by their Higgs phase VEVs,
$i m J_1$ and $im J_2$ respectively. At infinity, the profile
functions $\psi_1$ and $\psi_2$ approach unity, while $\psi_2$
vanishes at the origin so that the solution remains smooth. As we go around the
origin, $\Phi_1$ and $\Phi_2$ undergo a phase rotation generated by 
\beq 
Y_1= \frac{1}{3} {\rm Diag} (1,1,-2).
\eeq
The magnetic flux for the solution turns out to be proportional to
$Y_1$, which satisfies 
\beq
\exp(2 \pi i Y_1 )= {\rm Diag} (e^{ 2 \pi i /3},e^{ 2 \pi i /3},e^{2
  \pi i /3} ). 
\eeq
The gauge field is modified slightly from the $SU(2)$ case,
\beq 
A_x=\frac{-y}{r^2} \left(  (1-f(r)) Y_1  + g(r) \lambda \right), \,\qquad
A_y=\frac{x}{r^2}  \left(  (1-f(r)) Y_1  +g(r) \lambda \right), \eeq
where $\lambda=\frac{1}{2} { \rm Diag}(1,-1,0)$ and $g(r)$ is a new
profile function. The non-Abelian magnetic field then is 
\beq
F_{xy}= -{f'(r)\over r}\, Y_1 + {g'(r)\over r} \,\lambda.
\eeq
As in the $SU(2)$ example, we can evaluate the energy per unit length
for the ansatz to obtain the tension functional which can be varied to
yield the equations of motion,
\begin{eqnarray}
&&T = 2 \pi \int r \, dr \left( \frac{2 f'^2}{3 r^2} +\frac{g'^2}{2 r^2} +
\frac{1}{2}m^2 (4 \kappa_1'^2+3 \kappa_2'^2)+2 m^2 (\psi_1'^2+\psi_2'^2)+\right.\\\nonumber 
&&\left.+{1\over 2 r^2}m^2 \left(4 g^2 \psi _1^2+(g-2 f)^2 \psi _2^2\right) +
\frac{1}{2} m^4 \left(
  \frac{1}{2}(2\kappa_1-\kappa_2-2\psi_1^2)^2+\frac{1}{2}
(2\kappa_1+\kappa_2-2\psi_2^2)^2\right.\right.\\\nonumber
&&\left.\left. 2(\kappa_2+\psi_1^2-\psi_2^2)^2+
(2-2\kappa_1+3\kappa_2)^2\psi_1^2+(2-2\kappa_1-3\kappa_2)^2
\right)\right).
\end{eqnarray}
%(4 \psi _1^4+(2 -2 \eta _1+3 \eta _2)^2 \psi _1^2+4 \eta _1^2+3 \eta _2^2 ) + \right.  \]
%\[\left.+ \frac{1}{2} m^4 \left(4 \psi _2^4+\left(2 \eta _1+3 \eta
%_2-2\right)^2 \psi _2^2-2 \left(2 \eta _1+3 \eta _2\right) \psi  
%   _2^2+\psi _1^2 \left(-4 \psi _2^2-4 \eta _1+6 \eta _2\right)\right)
%\right) \, .\]  }
Once again the equations of motion following from this functional are 
consistent with the equations of motion of the full theory,
Eqs.(\ref{eqm}) and (\ref{eqm2}).

For $SU(3)$ gauge group there is also a $k=2$ vortex. However this is
follows from a $k \to N_c-k$ replacement in our $k=1$ solution. In
other words, the $k=2$ solution will be the identical to the above,
with the opposite flux (winding). 
\subsection{$N_c$=4}
The ${\mathbb Z}_{N_c}$ string solution for $SU(4)$ gauge group is
particularly interesting, as this is the first instance where we 
encounter a non-trivial multi-vortex solution, i.e. with 
winding $k>1$. We only need to consider the cases with $k=1, 2$ 
(the $k=3,4$ vortices are identical to $k=1,2$ respectively with
negative winding).
\\\\
{\underline{\it $k=1$ solution:}} The ansatz follows the general
pattern described earlier,
\begin{eqnarray}
&&\Phi_1= \frac{m i}{2 } \left(\begin{array}{cccc}
0  & \sqrt{3} \psi_1 & 0 & 0\\
\sqrt{3} \psi_1 & 0 & 2 \psi_2 & 0 \\
0 & 2  \psi_2  & 0 & \sqrt{3}  \psi_3 e^{i \varphi} \\
0 & 0 & \sqrt{3}  \psi_3 e^{-i \varphi} & 0\\
\end{array}\right), \\\nonumber\\\nonumber
&&\Phi_2=
\frac{m i}{2 } \left(\begin{array}{cccc}
0  & - i \sqrt{3} \psi_1 & 0 & 0\\
i \sqrt{3} \psi_1 & 0 & - i 2  \psi_2  & 0 \\
0 & i 2  \psi_2  & 0 & -i \sqrt{3}  \psi_3 e^{i \varphi }\\
0 & 0 & i \sqrt{3} \psi_3 e^{-i \varphi}& 0\\
\end{array}\right), \\\nonumber\\\nonumber
&&\Phi_3= \frac{m i}{2 } \left(\begin{array}{cccc}
3 \kappa_1 -2  \kappa_3 & 0 & 0 & 0\\
0 & \kappa_2 +  2  \kappa_3 & 0  & 0\\
0 & 0 & -\kappa_2  +  2 \kappa_3   &  0\\
0 & 0 & 0  & -3 \kappa_1 -   2 \kappa_3   \\
\end{array}\right).
\end{eqnarray}
The following expression is used for the gauge field:
\beq 
A_x=\frac{-y}{r^2} \left(  (1-f) Y_1  +\sum_{\ell=1}^2
g_\ell(r) \lambda_\ell \right), \,\qquad
A_y=\frac{x}{r^2}  \left(  (1-f) Y_1  +
\sum_{\ell=1}^2g_\ell(r) \lambda_\ell \right), 
\label{gfield}\eeq
where 
\beq
Y_1= \frac{1}{4} {\rm Diag} (1,1,1,-3).
\eeq
which yields $\exp(2 \pi i Y_1 )= {\rm Diag} (e^{ \pi i /2},e^{ \pi i /2},e^{\pi i
  /2},e^{\pi i /2})$. The non-Abelian flux carried by the vortex is
proportional to $Y_1$. The  
$g_\ell$'s are functions of $r$ vanishing both at $r=0$
and at $r \rightarrow \infty$, 
and $\lambda_\ell$ are a basis of diagonal matrices with satisfying,
\beq { \rm Tr} Y_1 \lambda_\ell=0, \qquad 
{\rm Tr} \lambda_i \lambda_\ell=\frac{1}{2} \delta_{i\ell}.  
\eeq
We choose
\beq 
\lambda_1=\frac{1}{\sqrt{12}} {\rm Diag}(1,1,-2,0), \qquad
\lambda_2=\frac{1}{2} { \rm Diag}(1,-1,0,0). 
\eeq
The string profile can then can be found by the minimization of the 
energy functional. We do not write the explicit form as it is
quite lengthy. The numerical solutions to the resulting equations of motion are 
shown in Figure 2.

\begin{figure}[h]
\begin{center}
%$\begin{array}{c@{\hspace{.2in}}c@{\hspace{.2in}}c} 
%\epsfxsize=1.6in
%\epsffile{nc4-1.eps} &
%    \epsfxsize=1.6in
%    \epsffile{nc4-2.eps} &
%    \epsfxsize=1.6in
%    \epsffile{nc4-3.eps}
%\end{array}$
\epsfig{file=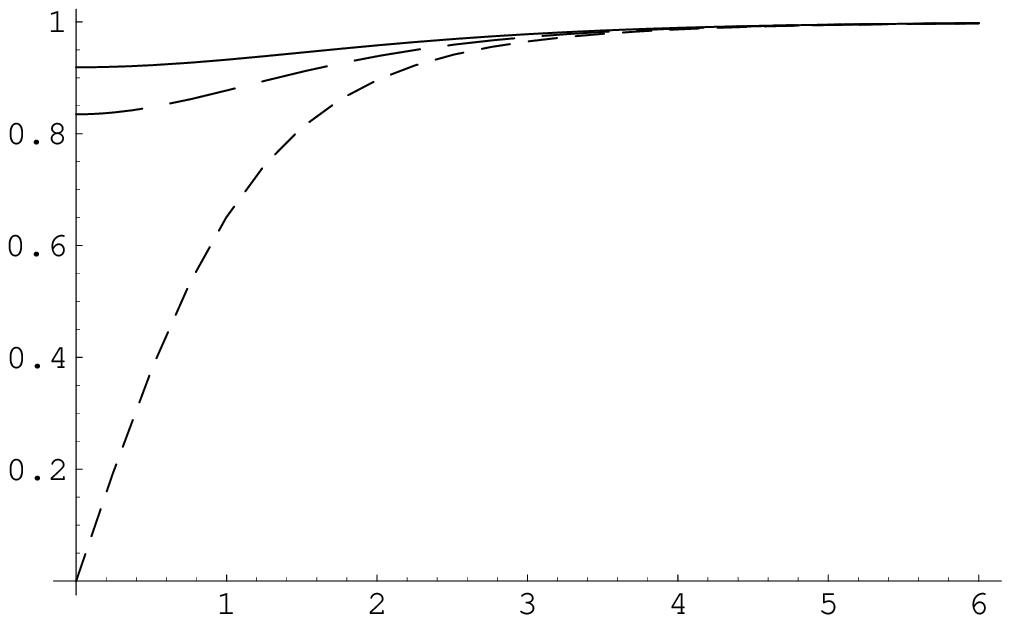, width =1.8in}\hspace{0.2 in}
\epsfig{file=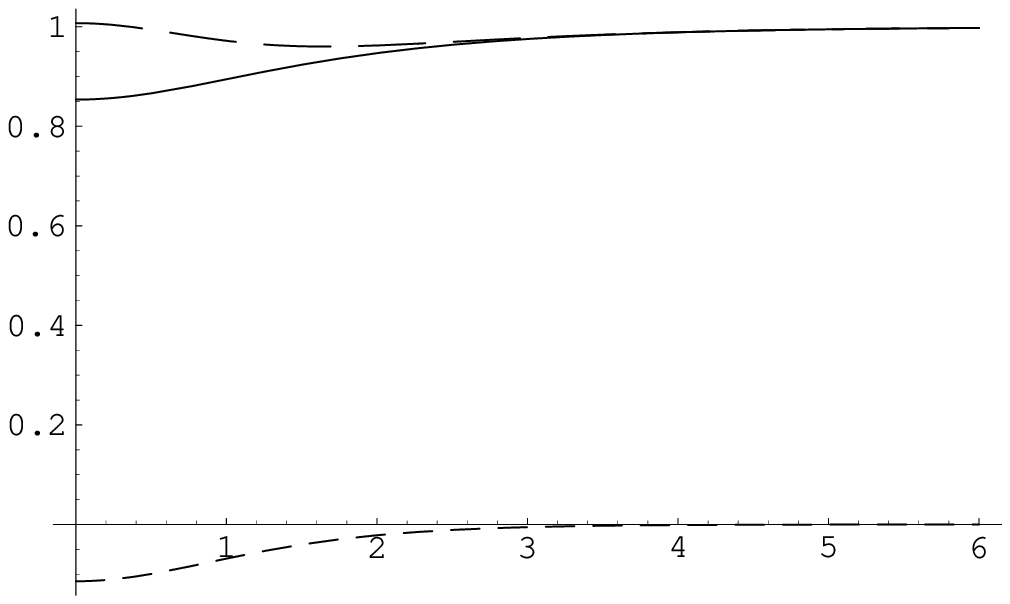, width=1.8in}\hspace{0.2 in}
\epsfig{file=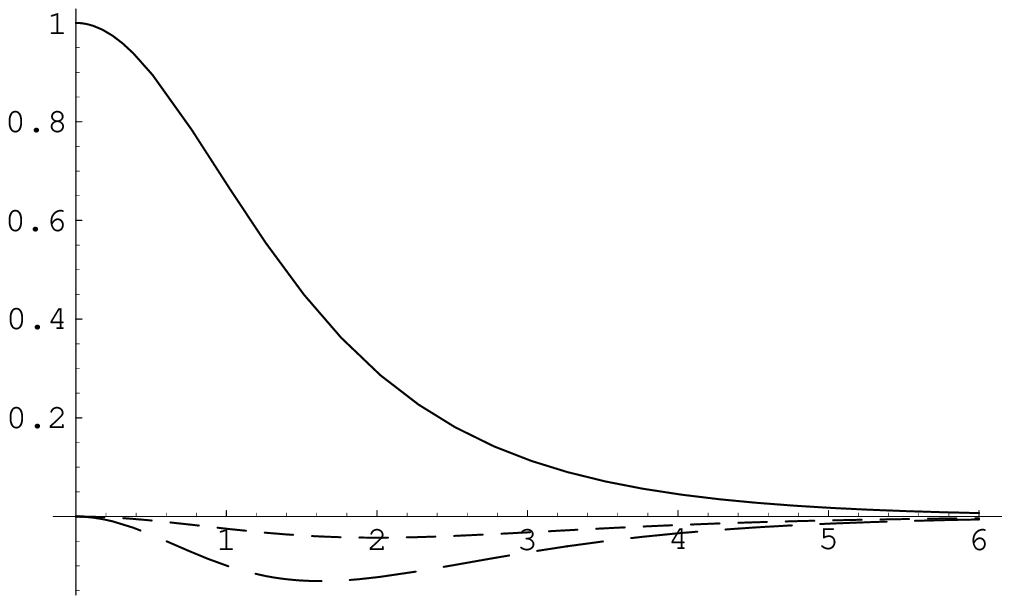, width=1.8in}
\end{center}
\noindent\small{\bf Figure 2:} The vortex profile for $N_c=4$. Left:
$\psi_{1}$ (solid), $\psi_{2}$ (long dashes), $\psi_{3}$ (short dashes).
Center: $\kappa_{1}$ (solid), $\kappa_{2}$ (long dashes), $\kappa_{3}$ (short dashes).
Right: $f$ (solid), $g_1$ (long dashes), $g_2$ (short dashes).
\label{nc4k1}
\end{figure}

Since $\psi_3$ vanishes at the origin and all the diagonal elements of
$\Phi_3$ remain non-zero at $r=0$, we infer that at the core of the
vortex solution, a $U(1)$ subgroup of the gauge symmetry is unbroken
and theory is in a Coulomb phase in that region. 
\\\\
{\underline{\it $k=2$ solution:}} We now turn to the $k=2$ vortex
solution. The relevant configuration for the scalars is now obtained 
by applying an $SU(4)$ rotation to the Higgs vacuum VEVs, generated by 
\beq
Y_2= {1\over 2}\,{\rm Diag}(1,1,-1,-1)
\eeq
with $\exp(2\pi i Y_2)= {\rm Diag}(-1,-1,-1,-1)$. The chromomagnetic
flux is also proportional to $Y_2$. The explicit ansatz is then,
\begin{eqnarray}
&&\Phi_1= \frac{m i}{2 } \left(\begin{array}{cccc}
0  & \sqrt{3} \psi_1 & 0 & 0\\
\sqrt{3} \psi_1 & 0 & 2 \psi_2  e^{i \varphi}& 0 \\
0 & 2  \psi_2  e^{-i \varphi}  & 0 & \sqrt{3}  \psi_3  \\
0 & 0 & \sqrt{3}  \psi_3 & 0\\
\end{array}\right),\\\nonumber\\\nonumber
&&\Phi_2=
\frac{m i}{2 } \left(\begin{array}{cccc}
0  & - i \sqrt{3} \psi_1 & 0 & 0\\
i \sqrt{3} \psi_1 & 0 & - i 2  \psi_2 e^{i \varphi } & 0 \\
0 & i 2  \psi_2  e^{-i \varphi} & 0 & -i \sqrt{3}  \psi_3 \\
0 & 0 & i \sqrt{3} \psi_3& 0\\
\end{array}\right), \\\nonumber\\\nonumber
&&\Phi_3= \frac{m i}{2 } \left(\begin{array}{cccc}
3 \kappa_1 -2 \kappa_3 & 0 & 0 & 0\\
0 & \kappa_2 + 2 \kappa_3 & 0  & 0\\
0 & 0 & -\kappa_2  + 2 \kappa_3  &  0\\
0 & 0 & 0  & -3 \kappa_1 - 2 \kappa_3  \\
\end{array}\right).
\end{eqnarray}
The gauge fields are still given by Eq. \ref{gfield},
with $Y_1$ replaced by $Y_2$ and
\beq
 \lambda_1=\frac{1}{2}{ \rm Diag}(1,-1,0,0), \,\qquad
\lambda_2=\frac{1}{2}{\rm Diag}(0,0,1,-1). \eeq
As before the vortex profiles can be found numerically and the results
are shown in  Fig. 3.

\begin{figure}[h]
\begin{center}
%$\begin{array}{c@{\hspace{.2in}}c@{\hspace{.2in}}c} \epsfxsize=1.6in
%\epsffile{nc4k2-1.eps} &
%    \epsfxsize=1.6in
%    \epsffile{nc4k2-2.eps} &
%    \epsfxsize=1.6in
%    \epsffile{nc4k2-3.eps}
%\end{array}$
\epsfig{file=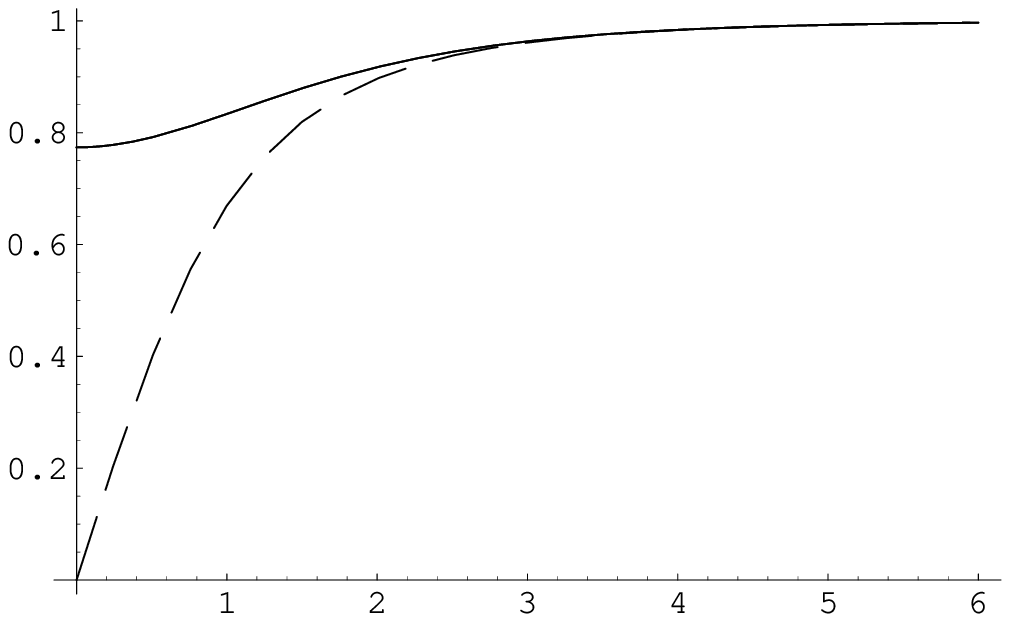, width =1.8in}\hspace{0.2 in}
\epsfig{file=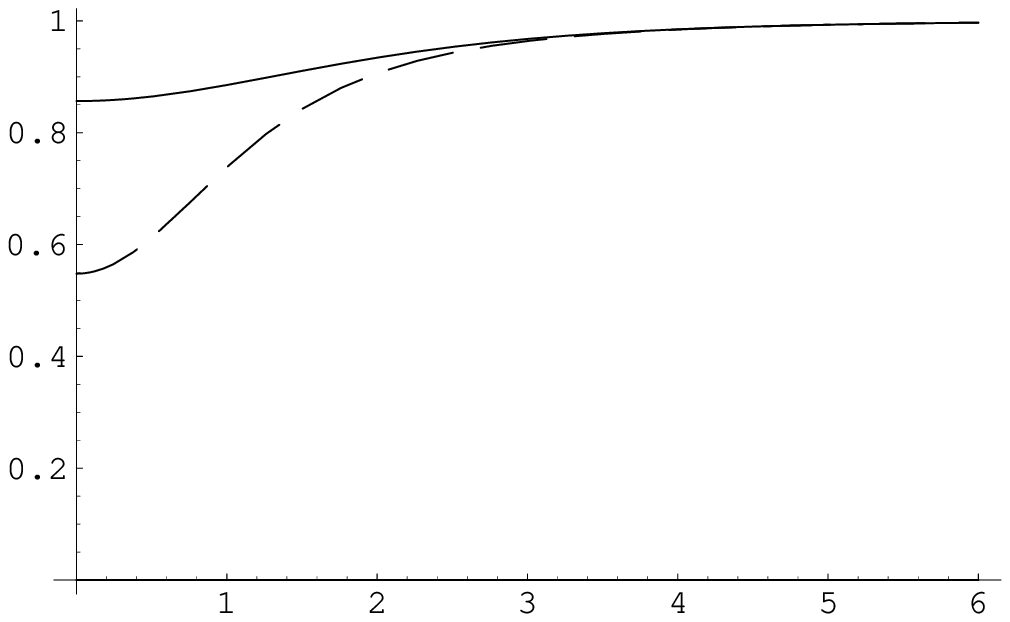, width=1.8in}\hspace{0.2 in}
\epsfig{file=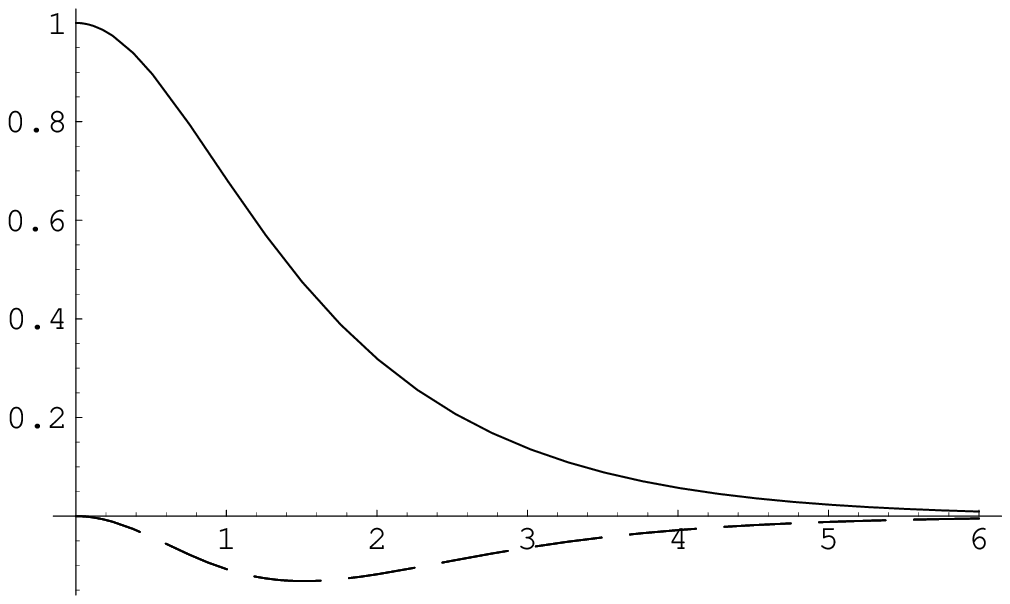, width=1.8in}
\end{center}
\noindent\small{\bf Figure 3:} Profiles for $N_c=4$ and $k=2$.
Left: $\psi_{1}=\psi_{3}$ (solid), $\psi_{2}$ (long dashes).
 Center: $\kappa_{1}$ (solid), $\kappa_{2}$ (long dashes), $\kappa_3=0$.
  Left: $f$ (solid), $g_1=g_2$ (long dashes).
\label{nc4k2}
\end{figure}

This solution provides a confirmation of the general picture of these
non-Abelian vortices in the Higgs vacuum of ${\cal N}=1^*$
theory. They all have an unbroken $U(1)$ gauge group at their core, while
approaching a totally Higgsed phase in the exterior. This is, of
course, consistent with the premise that the Higgs vacuum and its
excitations should have a semiclassical description. Another general
feature is that the non-Abelian strings break the $SO(3)_{C+F}$
symmetry group to a $U(1)$ subgroup. The moduli space of ${\mathbb
  Z}_{N_c}$ string solutions is therefore isomorphic to ${\mathbb
  {CP}}^1\simeq SU(2)_{C+F}/U(1)$ for all $k$ and $N_c$. We will also
confirm this feature of the theory in its large $N_c$ string dual.

The generalization of the vortex ansatz to arbitrary $N_c$ proceeds in
a straightforward fashion and requires introducing $3(N_c-1)$ profile functions
 ($\psi_i, \kappa_i, f, g_i$). In the absence of any obvious
 analytical simplifications, we will not pursue this direction further
 in this paper.

\subsection{$k$-string tensions }

The study of non-Abelian $k$-string tensions is a topic of great
interest and is particularly so in the present context. The
non-Abelian vortices of the Higgs vacuum at weak coupling $g_{\rm YM}\ll
1$ are mapped by S-duality of ${\cal N}=1^*$ theory to confining
strings at strong coupling $g_{\rm YM}\gg 1$. With the explicit ans\"atze
at hand for general $N_c$ and $k$, we can compute their tensions,
albeit numerically. We will then compare these results with the known
tensions in a different parametric regime for ${\cal N}=1^*$ theory wherein the
vortex strings are almost BPS. It should be pointed out that when
$m_1=m_2=m_3 =m$, the strings are {\em far from} BPS. Nevertheless we will
see that the numerical values of the tensions approach the BPS values
as $N_c$ is increased.

The tension of semiclassical non-Abelian strings in the Higgs vacuum
of ${\cal N}=1^*$ theory has been 
discussed in \cite{mamayung,kneipp}, in the limit 
\beq m_1=m_2=m,\quad{\rm and}\quad  m_3<<m.
\eeq
In this limit the vortex becomes an almost BPS object, due to the fact that
in the limit $m_3/m \rightarrow 0$, $\mathcal{N}=2$ supersymmetry is
restored. The theory may then be viewed as softly broken ${\cal
  N}=2^*$ theory. The ${\cal N}=2^*$ theory, with $m_1=m_2=m$ and $m_3=0$,
is realized in the Coulomb phase due to $\Phi_3$ obtaining a
VEV. Adding a mass $m_3$ for $\Phi_3$ at the appropriate point on the
Coulomb branch moduli space results in complete Higgsing of the theory
due to electric degrees of freedom becoming light and condensing. 
The profile functions of the vortices in this limit are simpler,
because it is consistent to take the profiles $\kappa_j(r)$
(equivalently, $\Phi_3$) as constant. 
 For a BPS vortex the tension is exactly
proportional to the field condensates:
\beq T_{N_c,k}^{BPS}= 2 \pi \, \frac{m m_3}{g_{\rm YM}^2} \, k (N_c-k)
. \label{tbps}\eeq 
This behaviour is the so-called `Casimir
scaling'  of $k$-string tensions.

The case we have focussed attention on this paper is far from the BPS
limit with,
\beq m_1=m_2=m_3=m.\eeq
Using our ansatz above we have numerically evaluated the vortex tension
functional  $T_{N_c, k}$ for $2\leq N_c \leq 6$ and the results are in
Table~\ref{tensii}. In this table we have presented the ratio of 
the tensions for the  $SO(3)_{C+F}$ symmetric theory to the BPS
formula (\ref{tbps}) extrapolated to $m_3=m$.
\begin{table}[h]     
\begin{center}    
\begin{tabular}  {|l|l|l|l|l|l|} 
\hline   $N_c$        & $2$ & $3$     & $4$ &   $5$ & $6$ \\ \hline
$k=1$  &$0.894$&$ 0.926$&$0.943$ &$ 0.954$&$ 0.961$ \\ \hline      
$k=2$  &   &              &$0.944$ &$ 0.954$& $0.962$   \\ \hline  
$k=3$  &   &              &          &      & $0.962$  \\ \hline     
\end{tabular}   
\caption{\footnotesize 
Values of $T_{N_c,k}/T_{N_c,k}^{\rm BPS}$ for $2\leq N_c \leq 6$ and different $k$.
 }     
\label{tensii}  
\end{center}
\end{table}

The main conclusion that we can draw from this numerical data is 
that for large $N_c$
the $k$-string tension $T_{N_c,k}$ in the theory with $m_3=m$
quickly approaches the BPS tension formula given by 
Eq.~(\ref{tbps}). There does not appear to be an 
obvious explanation for this result.
We also note also that for fixed $N_c$ the ratios in the table are,  
to a very good approximation,  independent of $k$.

The numerical results for string tension ratios
$T_{N_c,k+1}/T_{N_c,k}$ are also rather striking.
For $N_c=4$ we find the following numerical result,
\beq
\frac{T_{N_c=4,\,k=2}}{T_{N_c=4,\,k=1} }=1.334 
\eeq
while the prediction from  Casimir scaling is $4/3$.
For $N_c=5$ we find 
\beq
\frac{T_{N_c=5,\,k=2}}{T_{N_c=5,\,k=1} }=1.501 
\eeq
while the Casimir scaling prediction is $3/2$. Finally, for 
$N_c=6$:
\beq
\frac{T_{N_c,\, k=2}}{T_{N_c=6,\,k=1} }=1.6008, \qquad  
\frac{T_{N_c=6,\,k=3}}{T_{N_c=6,\,k=1}
}=1.801, 
\eeq
while the Casimir scaling values are $8/5$ and $9/5$.

The numerical results above are striking in that 
the tension is not a BPS protected quantity, so the accuracy of the Casimir
scaling law is better than what we could expect.
The Casimir scaling law is only exact in the limit $m_3<<m$,
but evidently it is still an extremely good approximation also for $m = m_3$
 for the cases $N_c=4,5,6$ which have been studied numerically. This
 suggests that in the large $N_c$ theory, the $k$-string tensions likely
 obey a Casimir scaling law in the ${\cal N}=1^*$ theory. This can
 be best understood by investigating the known large-$N_c$ string dual
 of the ${\cal N}=1^*$ theory \cite{ps}, which we will do in Section 5.

\section{Effective world-sheet theory} \label{effective}

In this section we round off our field theoretic analysis with the
 construction of the (classical) world-sheet theory of
$k$-strings in the Higgs vacuum. For $SU(2)$ gauge group this was
already done in \cite{mamayung}. Below we will extend this to
$SU(N_c)$ gauge group and general $k$. We will also present a new
ingredient, namely the effect on the worldsheet sigma model, 
of a non-zero $\theta_{3+1}$ angle in the ${\cal N}=1^*$ Yang-Mills theory.
 
The general class of vortex solutions presented above have the
property that they are invariant only under a $U(1)$ subgroup of
the colour-flavour locked $SO(3)_{C+F}$ transformations. This unbroken
$U(1)$ is a rotation acting on the pair $(\Phi_1, \Phi_2)$ which can
be undone by a gauge transformation. The moduli space of inequivalent
solutions is thus ${\mathbb {CP}}^1\simeq SU(2)_{C+F}/U(1)$. The
associated moduli correspond to the orientational modes of the
magnetic flux in the string solution.
 
The low-lying excitations of the worldsheet theory of the vortex will
 involve, apart from translational zero modes for the center of mass, 
the adiabatic dynamics of the orientational zero modes. For all $N_c$
and $k$  we see that this is a
nonsupersymmetric sigma model (as in the examples discussed in Refs.~\cite{mamayung,gsy}) 
with target space ${\mathbb {CP}}^1$,
along with a theta angle that is related in a special way to the four
 dimensional Yang-Mills theta  angle. 
The absence of supersymmetry makes the present situation
different from  BPS non-Abelian  vortex strings in ${\cal N}=2$ SQCD
\cite{HT, HT2,ABEKY,  Shifman:2004dr} and also different 
from the Heterotic vortex string discussed in Refs.~\cite{heterotic}.

\subsection{Kinetic term}

Let us consider a vortex oriented along the $z$ axis.
In order to obtain the effective world-sheet theory of the
orientational zero modes, we introduce an adiabatic $SO(3)_{C+F}$
rotation which depends on the world-sheet coordinates $(z,t)$ of the vortex
string. Doing so will turn these moduli (the global rotation
parameters) into worldsheet dependent fields. It is best to perform
these steps in singular gauge, i.e. where the scalar fields have no
winding at infinity and the flux is concentrated near the origin as in
\cite{ABEKY, Shifman:2004dr}.

Upon a (worldsheet dependent) colour-flavour locked  
rotation, the triplet of scalar fields transform in the following way:
\beq \vec{\Phi} \rightarrow U_F(z,t)\cdot
\left(W_C(z,t)\,\vec{\Phi}\,W_C^{\dagger}(z,t)\right),\eeq 
where the matrix $U_F$ acts in the three dimensional flavour space 
(\ref{flavour}) and
$W_C$ is a colour transformation (\ref{colour}) generated by the $N_c$ 
dimensional representation of $SU(2)$ generators.
The gauge fields transform as:
\beq 
A_{x,y} \rightarrow W_C\, A_{x,y}\,W^{\dagger}_C \, .
\label{zt}
\eeq
A transformation dependent on $z$ and $t$, will of course also
generate components  of $A_\mu$ along the world-sheet
coordinates, as evident from the ordinary gauge transformation
$
A_{s}\rightarrow W_C\, A_s\,W_C^\dagger+iW_C\,\partial_s W_C^\dagger$.
To be consistent therefore, 
the full vortex solution will need to be modified and the radial
dependence of the new
components of the gauge field have to be solved for.
We will, using a natural axisymmetric ansatz for all
$A_\mu$, obtain the worldsheet effective theory.
This can be done along the lines 
of \cite{gsy} : $A_s$ are choosen in gauge space in such a way that
they are perpendicular to the $A_{x,y}$ 
and to the derivative $\partial_s (A_{x,y})$. This is 

Let us discuss, for simplicity, the $N_c=2$ $k=1$ case (already studied in
\cite{mamayung}) and subsequently generalize. In singular gauge
\begin{eqnarray}
&&\Phi_1= \frac{i m}{2 }\psi_1(r) 
\left(\begin{array}{cc}
0  & \,1  \\
1   & \,0 \\
\end{array}\right), \,\quad
 \Phi_2=
\frac{i m }{2 } \psi_1(r)\left(\begin{array}{cc}
0  & -i   \\
i  & 0  \\
\end{array}\right),\\ \label{sing}
&&\Phi_3= \frac{i m }{2 }\kappa_1 \left(\begin{array}{cc}
1  & 0 \\
0 & -1 \\
\end{array}\right) \, ,  \,\quad
 A_x=\frac{y}{r^2} f(r) \frac{\sigma_3}{2}, \,\quad
A_y=\frac{-x}{r^2} f(r) \frac{\sigma_3}{2} \, .
\end{eqnarray}
Since the solution is symmetric under $U(1)_{C+F}$ rotations around
the $\Phi_3$ axis in flavour space, to generate a new solution we need
to act on it with one of the broken generators. If this action is
chosen to be global, we simply obtain a new solution with the the same
tension as the old one, but with the non-Abelian flux pointing in a
different direction in colour+flavour space. Let us therefore consider  
a $(z,t)$ dependent rotation around the $\Phi_2$ axis without loss of
generality. We will use the following worldsheet
dependent colour and flavour transformations:
\beq  W_C \,\sigma_3\, W^\dagger_C = \vec n(z,t)\cdot \vec \sigma\, . \eeq
Then the following ans\"atze can be used for the gauge field components 
$A_s$ with $s=z,t$: 
\beq A_s = -\left( \vec n \times \partial_s \vec n\right)^a {\sigma^a\over 2}\rho(r) \, .  \label{pagliaccio} \eeq

The gauge orientations for $A_s$ are dictated by the requirement that they be
orthogonal in colour space, to both $A_{x,y}$ and $\partial_s A_{x,y}$
(after the world-sheet dependent transformation). 
We substitute the expression for the gauge fields (\ref{pagliaccio}) along with the 
world-sheet dependendent transformation (\ref{zt}) of the solution 
(\ref{sing}), into the action and obtain
\beq
S_{1+1}= \int dz\,dt B_{2,1}(\partial_s \vec n)^2\,,
\eeq
where 
\begin{eqnarray} 
&&B_{2,1}= \\\nonumber
&&\frac{1}{g^2_{\rm YM}}
\int_0^\infty 
dr \, 2 \pi r \, \left( \frac{\rho'^2}{2}+\frac{(\rho-1)^2 f^2}{2
    r^2}
+\frac{m^2}{2}( 2 (\kappa_1-\psi_1)^2 (1-\rho) +(\kappa_1^2+\psi_1^2)
\rho^2 ) \right) \,. 
\end{eqnarray}
It is worth noting that this kinetic term for the world-sheet moduli
is generated only by gauge kinetic terms depending on $F_{s\mu}$
and scalar kinetic terms $\sim |D_s\Phi_{1,2}|^2$ of  
the four dimensional gauge theory. All other terms including 
the scalar potential of the gauge
theory are invariant under the combined colour-flavour rotations. 
In order for the sigma model coupling to be finite, 
we need to impose the boundary conditions 
\beq
\rho(0)=1\,,\qquad \rho(r\to \infty)=0.
\eeq
The Euler-Lagrange equations for $\rho(r)$ and other profile functions 
can be solved numerically to yield 
the kinetic term for the world-sheet moduli fields. The result, shown in Table
2 for $SU(2)$: $B_{2,1}=0.39 \,(2\pi/g^2_{\rm YM})$ matches with that of
\cite{mamayung}. 

% It is an easy excercise to follow the above steps for
% a general $SU(2)/U(1)$ transformation with both $z$ and $t$ dependence to get,
% \beq S_{1+1}= \int dz\,dt B_{2,1}(\partial_s \vec n)^2\,, \eeq
% with $\Tr(W_C\partial_s W_c^\dagger \sigma_3)=0$. Crucially, this general
% approach requires the general ansatz for $A_z$ and $A_t$, 
% \beq  A_s = \left( \vec n \times \partial_s \vec n\right)^a {\sigma^a\over 2}\rho(r).  \eeq

The generalization of the above arguments to the $SU(N_c)$ case, 
for each of the stable $k$-vortices, is actually straightforward.
Now the relevant colour transformations are generated by the $N_c$
dimensional representation of $SU(2)$ generators. To get the
normalization of the kinetic term of the resulting ${\mathbb{CP}}^1$
model,  we again  consider just a rotation around the
$\Phi_2$ axis so that,
\beq  W_C \, J_3 \, W^\dagger_C = \vec n(z,t)\cdot \vec J \, . \label{sucettes} \eeq
Then the following ansatz can be used for the gauge fields along the
worldsheet
\beq A_s = -\left( \vec n \times \partial_s \vec n\right)^a  \, J^a \rho(r) \, .  \label{pagliaccio2} \eeq
If we insert these expressions into the gauge theory 
action, the following term in the vortex  effective theory can be found,
\beq S_{1+1} = \int dz \, dt \,  \left(  B_{N_c,k} (\partial_s \vec{n})^2 
  \right).
 \eeq
The general formula for $B_{N_c,k}$ is complicated and we can only
evaluate it on a case by case basis, numerically.

For example, for $N_c=3$, $k=1$ the explicit expression is,
\begin{eqnarray} 
&&B_{3,1} = \\\nonumber
&& \frac{1}{g^2_{\rm YM}}
\int dr \, 2 \pi r \, \left( 2  \rho'^2+ \frac{\left(4 f^2-4 g f+5
      g^2\right) (\rho -1)^2}{4 r^2}+
\frac{1}{2} m^2 \left(4 \left(\rho ^2-2 \rho +2\right) \kappa
  _1^2+\right.\right.\\\nonumber
&&\left.\left.+8 (\rho -1) \left(\psi _1+\psi _2\right) \kappa _1+
3 \left(3 \rho ^2-6 \rho +4\right) \kappa_2^2 -
12 (\rho -1) \kappa _2 \left(\psi _1-\psi _2\right)\right.\right.\\\nonumber
&&\left.\left.+2 \left(-4 \psi _1
  \psi _2 (\rho -1)^2+ \left(3 \rho ^2-6 \rho +4\right) \psi _1^2+\left(3 \rho ^2-6
   \rho +4\right) \psi _2^2\right)\right)\right)\, . 
\end{eqnarray}
The coefficient is then determined numerically by variation of the
action functional with the boundary conditions $\rho(r=0)=1$ and $\rho(r\rightarrow\infty)=0$.
In Table \ref{betas} are shown the numerical values for $B_{N_c,k}$
for  $2\leq N_c \leq 6$.
\begin{table}[h]     
\begin{center}    
\begin{tabular}  {|l|l|l|l|l|l|} 
\hline   $N_c$        & $2$ & $3$     & $4$ &   $5$ & $6$ \\ \hline
$\frac{g^2_{YM}}{2 \pi} B_{N,1}$  &$0.390$&$ 1.181$&$ 2.343$ &$3.867$&$ 5.847$ \\ \hline      
$\frac{g^2_{YM}}{2 \pi} B_{N,2}$  &   &              &$ 2.696$ &$4.344$& $6.710$  \\ \hline  
$\frac{g^2_{YM}}{2 \pi} B_{N,3}$  &   &              &          &      & $6.888$  \\ \hline     
\end{tabular}   
\caption{\footnotesize Some numerical results for the classical kinetic term for the $k$-vortex.
For $N_c=2$ there is agreement with the value computed in Ref.~\cite{mamayung}. }     
\label{betas}  
\end{center}
\end{table}

\subsection{World-sheet Theta Angle}

Whenever 
the Yang-Mills theta angle $\theta_{3+1}$ is non-vanishing, a new
ingredient appears in the vortex world-sheet theory. 
The Yang-Mills theta angle feeds into the world-sheet theory as a topological
term for the ${\mathbb {CP}}^1$ sigma model. The coefficient of this
topological term, the world-sheet theta angle denoted as
$\theta_{1+1}$, plays a crucial role in the ensuing world-sheet
dynamics. In particular, the IR dynamics is strongly theta-dependent
\cite{gsy,zam1,zam2,zam3}.

We begin by demonstrating the mechanism of generation of the
world-sheet theta angle for $SU(2)$ gauge group. In this case the
steps involved and the result are rather similar to 
\cite{gsy}. The relevant terms can be obtained by a colour-flavour
transformation that depends on both $z$ and $t$. 
 It will be sufficient to consider the following $(z,t)$
dependent colour-flavour rotation of the vortex fields, 
\beq 
U_F=\exp(J_2 \alpha(z)).\exp(J_1 \beta(t))
 \, , \qquad W_C=\exp \left(i \frac{\sigma_2}{2} \alpha(z)\right).
 \exp \left(i \frac{\sigma_1}{2} \beta(t)\right)\, 
.\eeq
The time and space dependent rotation will generate gauge field
components $A_z$ and $A_t$. These will have to be chosen normal, in
colour space, to $A_{x,y}$ and their derivatives. Using the appropriate
ansatz,
$A_s = -(\vec n\times \partial_s \vec n).(\vec\sigma/2) \rho(r)$, we obtain
\begin{eqnarray}
&&A_z=  (\frac{\sigma_1}{4} \sin \alpha \sin 2 \beta+
\frac{\sigma_2}{2} \cos^2 \beta-\frac{\sigma_3}{4} \cos \alpha \sin 2
\beta )   \,\rho(r)\, \alpha'(z)\, ,
\\\nonumber
&&A_t=(\frac{\sigma_1}{2} \cos \alpha +\frac{\sigma_3}{2} \sin \alpha)
\,\rho(r)\, \dot\beta(t) \, .
\end{eqnarray}
Introducing the new world-sheet variations into the space-time action
action, the theta dependent topological term in the Yang-Mills action
then gives rise to a topological term on the world-sheet of the vortex
\beq 
{S}^{\theta}_{1+1}= \frac{\theta_{3+1}}{32 \pi^2}\int d^4x\, F^a_{\mu \nu}
\tilde{F}^a_{\mu \nu}=
\frac{\theta_{3+1}}{16 \pi^2} \int dz \, dt \,C \alpha'(z)
\dot\beta(t) \cos \beta 
\, ,\eeq
where 
\beq C=\int_0^\infty \,2 \pi \,dr\, r \,\frac{1}{2 r}\frac{d
  }{dr}\left (\rho^2-2 \rho )f \right)=\pi \,.
\eeq
Note that $C$ is obtained by integrating a total derivative and only
depends on the values of the profile functions at zero and infinity, namely
$\rho(0)=f(0)=1$ and $\rho(\infty)=f(\infty)=0$. Written more
covariantly, this  
leads to the following interaction in the vortex effective action,
\beq {\cal L}^\theta_{1+1}= -\frac{\theta_{3+1}}{8 \pi}
\epsilon^{s, r}\, \epsilon^{a b c}\, n^a \partial_s n^b \partial_{r} n^c
\qquad s,r = (t,z). \eeq
This is very similar to the case discussed in Ref.~\cite{gsy}
and relates the theta angle of the $\mathbb{CP}^1$ model to the
Yang-Mills theta angle as
\beq
\theta_{1+1}=\theta_{3+1}\,\qquad {\rm for}\quad SU(2).
\eeq
The general result for arbitrary $N_c$ and $k$ is more illuminating
than the $SU(2)$ theory. In particular, in the general case the
world-sheet theta angle is not equal to the space-time theta angle;
the two are related and this relation depends both on $N_c$ and $k$. 
We may consider the most general colour-flavour
rotated gauge field configurations in Eqs. (\ref{sucettes}) and (\ref{pagliaccio2}) and
evaluate the topological term on these to produce
\beq
{\cal L}^\theta_{1+1} = -\, C_{N_c,k}\;{\theta_{3+1}\over 8\pi^2}
\,\epsilon^{s r}\, 
\epsilon^{a b c}\, n^a \partial_s n^b \partial_{r} n^c.
\eeq
%It is howeveWe can generalization to arbitrary $(N,k)$ is straightforward:
%\beq U=\exp(T_2 \alpha(z)).\exp(T_1 \beta(t))
% \, , \qquad W=\exp (i S_2 \alpha(z)).
% \exp (i S_1 \beta(t))\, .\eeq
%with the ansatz
%\beq A_z=  (\frac{S_1}{2} \sin \alpha \sin 2 \beta+
%S_2 \cos^2 \beta-\frac{S_3}{2} \cos \alpha \sin 2 \beta )   \rho \alpha'\, ,\eeq
%\[ A_t=(S_1 \cos \alpha +  S_3  \sin \alpha) \rho \beta' \, .\]
The proportionality constant $C_{N_c,k}$ is again given by the
integral of a total derivative,
\begin{eqnarray}
C_{(N_c,k)}=\int_0^\infty 2 \pi  \, dr {d\over d r} \left\{(\rho^2-2\rho)
\left( f(r)\Tr (Y_{N_c,k} J_3) 
%+{g_k}(r) \Tr (\lambda_k J_3)
\right) \right\}
= \pi \,k (N_c-k).
\end{eqnarray}
This means that for the $k$-vortex, the theta angle of the world-sheet
sigma model is determined by $\theta_{3+1}$ as
\beq
\theta_{1+1} = k (N_c-k)\theta_{3+1}.
\eeq
%\rho (2-\rho) +\right. \eeq
%\[  \left.
%+  2 \rho' (1-\rho) \frac{\Tr (Y_{N,k}  S_3) f + \Tr (\lambda_k S_3) g_k }{r}
%  \right\} = \pi (N-k)  k \, .\]
So the long-wavelength fluctuations of the world-sheet theory of
the $\mathbb{Z}_{N_c}$ flux tube carrying $k$ units of magnetic flux,
are governed by the effective action
\beq  S_{1+1} = \int dz \, dt \,  
\left(  B_{N_c,k} (\partial_s \vec{n})^2 
 -k (N_c-k)\frac{\theta_{3+1}}{8 \pi}
\epsilon^{s r}\,\epsilon^{a b c} \,n^a \partial_s n^b \partial_{r} n^c  \right) \, ,\eeq
%where
%\beq \theta^{N,k}_{1+1}=k (N-k) \theta_{3+1} \, .\eeq
The effective theta angle is an integer multiple
of the four dimensional one and is thus guaranteed 
to respect the invariance of the Higgs vacuum under 
$\theta_{3+1} \rightarrow \theta_{3+1}+2 \pi$.

\subsubsection{Dynamics on the vortex world-sheet}
We have seen that the effective long-wavelength 
dynamics of the $k$-vortices in the Higgs vacuum
of ${\cal N}=1^*$ theory with $SU(N_c)$ gauge group (and with three
equal masses), is given by a
$\mathbb{CP}_1$ model for all $N_c$ and $k$. The four
dimensional theory being ${\cal N}=1$ supersymmetric, the
vortices are non-BPS and the effective world-sheet theory is
non-supersymmetric. Thus there are no fermionic super-orientational
zero modes. The resulting world-sheet dynamics is  different
from that of BPS vortex strings in ${\cal N}=2$ SQCD \cite{HT,ABEKY}
for example.

It is well-known that the value of the theta angle has a strong effect
on the IR dynamics of the $\mathbb{CP}^1$ model
\cite{witten,zam1,zam2,zam3,controzzi,gsy}. First of all the
$\mathbb{CP}^1$ model is asymptotically free and so is a strongly
coupled theory. This is interesting: the four dimensional field theory
is weakly coupled, but the dynamics on the vortex is highly quantum.
When $\theta_{1+1}=0$ and $\theta_{1+1}=\pi$, the model is exactly
solvable. Specifically, the spectrum at $\theta_{1+1}=0$ is known to
consist of a single massive $SO(3)$ triplet with an exact S-matrix
\cite{zam1} and the theory has a mass gap. This picture continues to
be valid for generic non-zero values of $\theta_{1+1}$. When
$\theta_{1+1}$ hits $\pi$, however, something drastic happens. The
theory has massless excitations and  flows to a $c=1$
conformal fixed point \cite{affleckhaldane, shankarread, zam2}
described by the $SU(2)$ Wess-Zumino-Witten model at level $k=1$. The
spectrum now consists of massless $SU(2)$ doublets. The picture,
therefore, is that at generic $\theta_{1+1}$, the doublets are
confined and bound into meson-like excitations, transforming as a
triplet of $SO(3)$. The singlet state, not having a conserved quantum
number, is unstable. It is possible to analyze the spectrum in the
vicinity of $\theta_{1+1}=\pi$ \cite{gsy,controzzi} and can be
interpreted as consisting of ``kink-anti-kink'' bound states. The
string tension between these kinks and anti-kinks (the $SU(2)$
doublets) vanishes as the vacuum angle approaches $\pi$.

The existence of the non-trivial dynamics near $\theta_{1+1}=\pi$ begs
the question: how is this reflected in the physics of the four
dimensional gauge theory? The
situation is particularly intriguing, since nothing obviously drastic happens in
the gauge theory when $\theta_{3+1} =\pi/k(N_c-k)$. This merits deeper
study, but one obvious possibility is that this concerns the spectrum
of confined monopole-dyon states in the Higgs phase. The doublets
(kinks) are likely to be the bound states of monopoles with the
vortex. These monopole-dyon states exist as massive 't Hooft-Polyakov
monopoles in the ${\cal N}=2^*$ theory in the Coulomb phase.
As $\theta_{3+1}$ is dialled, the spectrum of these massive states
undergoes a rearrangement and can lead to level crossing between
certain mutually non-local states (e.g. the (0,1) monopole and the
(1,1) dyon for $SU(2)$). It is possible that the special
values of $\theta_{3+1}$ may be the points at which such massive,
mutually non-local states become degenerate. If both these states
happen to get confined upon breaking the supersymmetry to ${\cal
  N}=1^*$, then they can appear bound to the magnetic flux tubes. The appearance of
such mutually non-local states simultaneously on the world-sheet, may
drive the sigma model to an interacting fixed point. We are merely
speculating at this stage, but clearly the issue deserves deeper
study.

\section{The Vortex in the String Dual}   \label{sugra}

The string theory dual of ${\cal N}=1^*$ theory was constructed by
Polchinski and Strassler \cite{ps} by considering an appropriate
deformation of Type IIB string theory on $AdS_5\times S^5$ background.
The undeformed $AdS_5 \times S^5$ background with $N_c$ units of
Ramond-Ramond five form flux is dual to the large $N_c$ limit of
$SU(N_c)$, ${\cal N}=4$ SYM. The relation between gauge theory and
string theory parameters is as follows. The string coupling $g_s$ and
the radius of curvature of AdS space are related to the gauge coupling
and the 't Hooft coupling respectively as 
\beq
4\pi g_s = g^2_{\rm YM}\,,\qquad {R_{\rm AdS}\over \sqrt{\alpha'}}= (4\pi g_s
N_c)^{1/4}\gg 1\,, \qquad C_0 ={\theta_{3+1}\over 2\pi}
\eeq
where $C_0$ is the Type IIB RR scalar.

The ${\cal N}=1^*$ mass deformation of the ${\cal N} =4$ theory is achieved by
switching on a non-normalizable mode for the three-form flux $G_3= F_3
- \tau H_3$, with $\tau = i/g_s + C_0/2\pi$, the unperturbed Type IIB
coupling. The Polchinski-Strassler dual geometry was obtained by
treating the $G_3$ flux as a perturbation and solving the Type IIB
supergravity equations of motion to linear order in this perturbation.
The rich infrared physics of ${\cal N}=1^*$ theory was captured in the
string dual using two central ingredients: the Myers dielectric effect
\cite{myers} and the action of $SL(2,{\mathbb Z})$ duality on the
vacua of the theory.

The classical description of the ${\cal N}=1^*$ vacua
\cite{ps}, 
shows that the scalars get noncommuting
expectation values describing fuzzy sphere configurations \cite{madore}.
 In the language of D-branes, this means that that $N_c$ D3-branes on which the parent
${\cal N}=4$ theory lives, acquire non-commuting positions, transverse to their
worldvolume. These transverse positions trace out fuzzy $S^2$'s and
the configuration can be reinterpreted as 5-branes wrapped
on concentric flux supported two-cycles carrying $N_c$ units of
D3-brane charge. The full large $N_c$, IIB string dual background interpolates 
between the ``near-shell'' geometry generated by the multiple 5-branes
and the asymptotically AdS solution towards the boundary of the
space. Different ${\cal N}=1^*$ vacua, with the theory realized in
different phases, are obtained by the action of the IIB $SL(2,{\mathbb
  Z})$ transformations on a given fivebrane configuration.

\subsection{Polchinski-Strassler Higgs Vacuum }

The Polchinski-Strassler description of each ${\cal N}=1^*$ vacuum
consists of an asymptotically AdS geometry with a $G_3$ flux turned
on, matching onto an interior geometry generated by $(c,d)$ 5-branes. 
For instance, classical vacua preserving an $SU(p)$ gauge
symmetry, where $p$ is a divisor of $N_c$ are described by $p$ coincident
D5 branes carrying a net D3 charge, provided
\beq
{q \over p} {1\over g_s}\gg 1\,,\qquad q = {N_c\over p}.
\eeq
When $p$ and $q$ are such that $p g_s/q\gg 1$, the vacuum with $SU(p)$
gauge symmetry is described by $q$ NS5 branes.

The Higgs vacuum is thus described by a single D5 brane carrying
net D3 charge, 
when 
\beq
{N_c\over g_s}\gg1
\eeq
while the confining vacuum is described by a single NS5 brane when
\beq
{N_c \,g_s}\gg 1\,.
\eeq
The former is an extremely weak condition in the large $N_c$ limit,
while the latter is the usual condition for the gauge theory to be
strongly coupled. The two vacua and these two 
conditions for the validity of the Polchinski-Strassler supergravity
description in the  ``far-from-shell'' region, are exchanged under
the S-duality, $g_s\leftrightarrow 1/g_s$. 

In the Higgs vacuum, with $m_1=m_2=m_3=m$,
the metric in the interior matches onto the
geometry generated by a D5-brane wrapped on an $S^2$ carrying $N_c$
units of D3-charge. The D3-brane worldvolume coordinates are 
$x^\mu$, $(\mu=0,1,2,3)$ wherein the field theory lives. The six
transverse directions are denoted as
\beq
w^i = x^{7,8,9}\,,\qquad {\rm and} \qquad y^i = x^{4,5,6}.
\eeq
The D3 branes spread out along the $w^i$ directions with $y^i=0$ and
the resulting 
D5-brane wraps a round sphere of radius
\beq
r_0 = \pi \alpha' m N_c.
\label{ro}
\eeq
The string frame metric   
of the Polchinski-Strassler solution \cite{ps}
corresponding to the Higgs vacuum (D5-brane) with equal masses for the
adjoint chiral multiplets and with 
$\theta_{3+1}=0$, is
\beq 
ds^2_{\rm string}=Z_x^{-1/2} \eta_{\mu \nu} dx^\mu dx^\nu+
Z_y ^{1/2} (dy^2+y^2 d \Omega_y^2 + dw^2)+
Z_\Omega^{1/2 } w^2 d \Omega_w^2,
\label{hmetric}
\eeq
where
\beq Z_x=Z_y=\frac{R^4_{\rm AdS}}{\rho_+^2 \rho_-^2}, \,\qquad
Z_\Omega=\frac{R^4_{\rm AdS}\,\rho_-^2}  {\rho_+^2  (\rho_-^2+\rho_c^2)^2}, 
\,\qquad \rho_{\pm}=\sqrt{ y^2+(w \pm r_0)^2 }. \eeq
The parameters in the metric are
\beq R^4_{\rm AdS}=4 \pi g_s N_c \alpha'^2, 
\qquad \rho_c=(\alpha' m) \sqrt{g_s N_c \pi} \, .
\eeq 
and $r_0$ is as defined in (\ref{ro}). The dilaton is non-constant,
approaching its asymptotic value $g_s$ as $w^i, y^i \to \infty$, but
vanishing close to the D5/D3 brane,
\beq e^{2 \Phi}=g^2_s \frac{\rho_-^2}{\rho_-^2+\rho_c^2},\,
\qquad 
C_0 =\theta_{3+1}=0.
\eeq
The Polchinski-Strassler (including the R-R and NS-NS potentials)
background has a manifest $SO(3)$ isometry acting on the $S^2$ 
in the $w$-plane. This naturally gets identified with the colour-flavour
locked $SO(3)_{C+F}$ symmetry of the Higgs phase of ${\cal N}=1^*$ theory.

The full solution above is  approximate, and is  
constructed by matching the $r\gg r_0$ limit with the ``near-shell''
solution for $r \approx r_0$. The 
``far-from-shell'' region is well approximated by the background
 generated by D3-brane charge density spread out on a spherical
 shell and is close to a Coulomb branch configuration. 
 In this regime the D3-brane charge density dominates over the D5
 charge density. The ``near-shell'' regime 
 is well described by the exact solution for a flat D5-brane with 
D3-brane charge \cite{AOSJ,BMM}. Here the effect of the D5-brane dominates.
The flat D5 solution and the matching used by Polchinski and Strassler
are reviewed in the Appendix.

An important feature of the Higgs vacuum geometry is that near the
D5-brane, supergravity ceases to be applicable. 
This is due to large transverse curvatures near the D5-brane. We will,
however, adopt a pragmatic approach and 
use the metric for our subsequent
analysis, with the aim of identifying certain aspects of the gauge
theory vortex dynamics that are robustly captured by the string dual.
The questions that we are interested in, will, perhaps surprisingly, turn out to be 
insensitive to the strongly curved parts of the geometry.

The main ingredient we will need from the region near 
the spherical D5 shell, is the NS-NS two-form potential (see the Appendix
for further details),
\beq 
B_2=- \frac{ \alpha' \pi N_c}{1+\rho_-^2/\rho_c^2}  \, \sin
\theta_w \, d\theta_w \wedge d \phi_w \, .
\label{flatb2}
\eeq 
Near the shell there are also the non-vanishing R-R potentials, $C_2$
and  $C_4$,  
which can be extracted from the flat D5 background discussed in Appendix.
The form of the R-R potentials at $\theta_{3+1}=0$ will not be relevant for the 
probe branes that we will study in this section. 

The Higgs vacuum solution at any non-zero theta angle for the gauge
theory follows upon acting on the solution with $\theta_{3+1}=0$
with an $SL(2,\mathbb{R})$ transformation in IIB supergravity. Under such
a transformation, $\tau\rightarrow \tau + {\theta_{3+1}\over {2\pi}}$, the 
5-brane remains a D5-brane, but the two R-R potentials $C_2$ and $C_0$
are shifted as,
\beq C_0 \rightarrow C_0 + \frac{\theta_{3+1}}{2 \pi}\, 
\qquad C_2 \rightarrow C_2+ \frac{\theta_{3+1}}{2 \pi} B_2 \,
. \label{thetarotaz} \eeq 
Thus at non zero $\theta_{3+1}$, the R-R two form $C_2$ acquires components
along the 
$d\theta_w \wedge d \phi_w$ direction which will be relevant for our probes.

\subsection{The Vortex as a D1-brane}
Magnetic flux tubes in the Higgs vacuum have a natural brane interpretation
as bound states of D1-branes with the D5-brane. Such D1-D5/D3 bound
states involving the single D5-brane in the Higgs vacuum are possible
due to the non-vanishing $B_2$ potential on the 5-brane, responsible for the
D3-charge. In this situation, the bound state is a semiclassical
instanton of the non-commutative field theory \cite{Nekrasov} on the 5-brane.

We first review the computation of the D-string tension, dual to a
$k=1$ vortex, in the Higgs vacuum geometry, first done in 
\cite{ps}. The two new ingredients in our analysis will be a
derivation of the world-sheet sigma model of the vortex 
and the effect of the Yang-Mills theta angle.

%If we restrict at $\vec{y}=0$, the metric reads:
%\beq \frac{w^2-r_0^2}{R^2} \eta_{\mu \nu} dx^\mu dx^\nu+
%\frac{R^2}{w^2-r_0^2} dw^2+ 
%\frac{R^2  (w-r_0)}{(w+r_0)((w-r_0)^2+\rho_c^2) } \, w^2 \, d
%\Omega_w^2 \, .\eeq 
Following \cite{ps}, we model a magnetic vortex as a D1-brane probe in
this geometry.  
The DBI action for the probe D1 brane in the geometry reads,
\beq S_{\rm DBI}=\frac{1}{2 \pi \alpha'} \int d^2 \xi \left\{ e^{-\Phi}
\sqrt{
 (- {\rm det} (G_{ab}+B_{ab}+2 \pi \alpha' F_{ab} ) ) } \right\}. \eeq
Let us consider a D1-brane oriented in the $x_0,x_1$ directions and
the embedding $(\xi_0, \xi_1)= (x_0,x_1)$, so that the pullback of the
metric onto the world-sheet is 
\beq G_{00}=-Z_x^{-1/2}+ \ldots, \,\qquad G_{11}=Z_x^{-1/2}+ \ldots \eeq
The dots correspond to terms involving fluctuations of the
string in the transverse $\vec y$ and $\vec w$ directions, proportional to
the derivatives $\partial_{0} (\vec{y},\vec{w})$, $\partial_{1}
(\vec{y},\vec{w})$. While the fluctuations
along $\vec y$ coordinates cannot be studied in supergravity due to
large curvatures, angular fluctuations in the $\vec w$ directions
will appear to be accessible.
The part of the DBI action which does not depend on
these derivatives and yields the effective tension  of the D1-brane is,
\begin{eqnarray}
&&S_{\rm DBI}=
\frac{1}{2 \pi \alpha'} \int d^2 x
(Z_x^{-1/2} e^{-\Phi})\\\nonumber
&&=\frac{1}{2 \pi \alpha'} \int d^2 x
\frac{\sqrt{y^2+(w+r_0)^2}  \sqrt{y^2+(w-r_0)^2+\rho_c^2 }  }
{g_s \,R^2_{\rm AdS}}. 
\end{eqnarray}
This is minimized when,
\beq 
w=\frac{r_0+\sqrt{r_0^2-2 \rho_c^2}}{2}\approx r_0 - \frac{\rho_c^2}{2 r_0},
\qquad  y=0. 
\eeq
Since $\rho_c\ll r_0$, the 
probe D1-brane sits at a relatively small distance $\delta w$ 
from the D5 shell,
\beq \delta w = \frac{\rho_c^2}{2 r_0} = \frac{g_s \alpha' m}{2} \, .\eeq
The dilaton and the functions that determine the metric at this point
evaluate to 
\beq e^{-\Phi} \approx 2 \sqrt{\frac{\pi N_c}{g^3_s} }, \qquad Z_x^{-1/2}
\approx \frac{\sqrt{g_s N_c \pi} \alpha' m^2}{2}, 
\qquad Z_\Omega^{1/2}= {1\over 2}\sqrt{ \frac{g_s}{(\pi N_c)^3  } }   
\frac{1}{\alpha' m^2} \,.\eeq
The curvature of the space transverse to the D5-brane , i.e., in
the radial $w$ and $\vec y$ directions can be seen to be substringy
at the value of $w$ giving the location of the D1-brane. However, as
we see below, this value for $w$ corresponds to a large radius sphere in 
the $w^{1,2,3}$ space. 

Continuing to use the metric (\ref{hmetric}), 
the tension of the D1-brane at this location is 
\beq T_{\rm D1} \approx \,\frac{2 r_0 \rho_c}{g_s R^2_{\rm AdS}} \,\frac{1}{2
  \pi \alpha'} 
= \,\frac{ N_c m^2}{2 g_s} \,= \,\frac{2 \pi N_c m^2}{g_{\rm YM}^2}
\, .\label{ts} \eeq
Remarkably, this formula matches the BPS formula (\ref{tbps}) for
$k=1$, in the Higgs vacuum which is only expected to work for softly
broken ${\cal N}=2^*$ theory. This is suggestive 
that the vortices at large $N_c$ (and $N_c/g_s\gg 1$) in the Higgs
vacuum become BPS objects. Equivalently, the confining strings at
large $N_c$ and $g_s N_c \gg 1$ (S-dualizing the Higgs vacuum) obey
the BPS
tension formula. More evidence in support of this possibility was offered in
\cite{ps}. This is also in agreement with our semiclassical results
for large $N_c$.

At its equilibrium position the D1-brane is transverse to a two-sphere 
at $y=0$ and $w=r_0 - g_s \alpha' m/2$, which is concentric with the 
dielectric 5-brane sphere. At this location the radius of
the transverse two-sphere is 
\beq
Z_\Omega^{1/4} w\approx {R_{\rm AdS}\over 2} =
{1\over 2}(4 \pi g_s N_c)^{1/4} \sqrt{\alpha'}.
\eeq
Clearly, this is large in string units in 
the supergravity limit. The D-string is pointlike on the transverse
two-sphere, resulting in a ${\mathbb {CP}^1}$ moduli space of vortex
solutions. The sphere is large in string units and we  
can allow for a slow, adiabatic
variation of the D1-brane position on the sphere, as a function of the
world-sheet coordinates. 
The polar coordinate of the D-string $\vec n_w \equiv (\theta_w,\phi_w)$,
corresponds to the vortex colour-flavour zero mode. 

Let us consider an arbitrary
dependence of $\vec n _w$ on the world-sheet coordinates $(x_0,x_1)$
and introduce this into the DBI action. 
Taking into account only 
the contribution of the pullback of the spacetime metric,
the following world-volume action results,
\beq 
S_{\rm DBI}=\int d^2 x \,\frac{e^{-\Phi}}{2 \pi \alpha'}\,
\sqrt{ -{\rm Det} \left(\begin{array}{cc}
-Z_x^{-1/2}+(\partial_0 \vec{n}_w)^2 w^2 Z_{\Omega}^{1/2}  &
(\partial_0 \vec{n}_w)\cdot (\partial_1 \vec{n}_w) w^2 Z_{\Omega}^{1/2}
\\\\
 (\partial_0 \vec{n}_w)\cdot(\partial_1 \vec{n}_w) w^2 Z_{\Omega}^{1/2} &
 Z_x^{-1/2}+(\partial_1 \vec{n}_w)^2 w^2 Z_{\Omega}^{1/2} \\ 
\end{array}\right) }
\eeq
In this formula we have actually omitted the pullback of the $B_{2}$
field, 
which only contributes to a four-derivative term 
in the vortex world-volume action that we neglect.
At the two-derivative level, we find
\beq S_{\rm DBI} \approx  T_{\rm D1}  \int d^2 x
\left( 1 +  (\partial_s \vec{n}_w)^2 \frac{w^2 Z_{\Omega}^{1/2}  Z_{x}^{1/2}}{2 } \right), \eeq 
where $T_{\rm D1}$ is the tension of the $k=1$ vortex in Eq.(\ref{ts}).
From this we get the ceofficient of the 
kinetic term of the $\mathbb{CP}^1$ sigma model,
\beq 
{\cal L}_{\rm kin}=\frac{N_c}{4 g_s} (\partial_s \vec{n}_w)^2 =\frac{ \pi
  N_c}{g_{\rm YM}^2} (\partial_s \vec{n}_w)^2 \,. 
\eeq
While it is interesting to perform the above formal manipulations, 
it is not clear that the classical coupling constant of the sigma
model is significant since the ${\mathbb{CP}^1}$ model
is asymptotically free and the coupling constant will run when the
sigma model is quantized. Secondly, the coupling can also get large
corrections due to stringy effects from the highly curved transverse
parts of the geometry. The coefficient of the topological term in
the ${\mathbb {CP}}^1$ model, on the other hand, has special
significance. We now turn to evaluating this from the D1-brane action.

\subsubsection{The Theta term}
To complete the supergravity analysis of the $k=1$ vortex, we will
now see how the Yang-Mills theta terms feeds into the world-sheet
sigma model. This feeding-in occurs through the Chern-Simons terms of
the D1-brane world-volume theory, 
\begin{eqnarray}
&&S_{\rm CS}=\\\nonumber
&&\frac{1}{2 \pi \alpha'} \int \,
 \left[ \exp(2 \pi \alpha' F_2+ B_2) \wedge \sum_q C_q  \right]=
\frac{1}{2 \pi \alpha'} \int \,
 \left[ C_2+ C_0 \, B_2  \right] \, \label{wzu}
\end{eqnarray}
We expect this term to have universal, robust features for two
reasons. First, the effect of a non-zero $\theta_{3+1}$ has to be such
that physics is periodic under shifts of the theta angle by
$2\pi$. This is particularly true in the Higgs vacuum which is 
actually invariant under such shifts, while the confining
vacuum gets mapped to an oblique-confining phase under the same operation. 
Furthermore, the Wess-Zumino term in the D-brane action is insensitive to the
background metric and may well capture the correct physics even in the
supergravity approximation.

Since the probe D1-brane is located relatively 
very close to the D5-brane shell, 
we need to use the expressions for $B_2, C_2 $ and  $C_0$
given in (\ref{thetarotaz}), (\ref{flatb2}) and 
(\ref{flatd5limit}) for the
``near-shell'' region.
There is a subtlety surrounding
the Wess-Zumino couplings of D-branes involving the pullback of the $B_2$
field (such as $C_0 \, \wedge B_2 $ in Eq. (\ref{wzu})), 
which has been discussed in the references \cite{Btricky}. The upshot
of this is that the contribution to Wess-Zumino terms from the
pullback of $B_2$ have to be omitted.
The term  $C_0 \wedge B_2 $ is an effective D(-1) `charge' 
for our probe D1, arising from the pullback of the $B$ field. 
In our case the contribution corresponds
to a D1 world-sheet wrapping an $S^2$ with $|\vec{w}| \approx r_0
-\rho_c^2/2r_0$, 
which is a homotopically trivial cycle. 
From the results of \cite{Btricky}, 
this term is cancelled by bulk contributions,
so we have to drop it from the present calculation.

Finally then, the only relevant term is the pullback of the components of
$C_2$ along the sphere at constant $w$ and $y=0$,
\beq
S_{CS} = {1\over 2\pi\alpha'}\int C_2 = {1\over 2\pi \alpha'} 
{\theta_{3+1}\over 2\pi} \int B_2 \big|_{\theta_{3+1}=0}.
\eeq
In the large $N_c$ supergravity limit, the magnitude of the near shell 
$B$ field (\ref{flatb2}) at the the radial position of the D1-brane, is equal 
to $\alpha' \pi N_c$.
The result is a
 two-derivative theta term for the effective $\mathbb{CP}^1$ sigma model,
\beq  
{\cal L}_\theta=\frac{\theta_{1+1}}{8 \pi}\,  
 \,\epsilon^{s r} \,\,\vec n_w \cdot 
\left( \partial_s \vec n_w \times \partial_{r} \vec n_w\right)\,,
\,\qquad \theta_{1+1}=  N_c \theta_{3+1} \,\qquad (s,r) = x^{0,1}\,. \eeq
This is consistent (at large $N_c$) with our semiclassical field
theory calculation done in the previous section. 
Note that if we were to keep the term proportional to $C_0 \wedge B_2 $,
we would find  
$\theta_{1+1}=  2 N_c \theta_{3+1} $, which would be inconsistent with
our semiclassical expectation.

\subsection{The $k$-vortex as a D3 brane}

Although we were able to reproduce the sigma model of the $k=1$ vortex
using the dual geometry, the tensions and the theta terms for
$k$-strings have a nontrivial dependence on $k$. It is not {\it a
  priori} clear that the Polchinski-Strassler geometry should be able to
reproduce these since the IR geometry for the Higgs vacuum will
receive large stringy corrections. However, as before, we will attempt
to identify the appropriate D-brane configuration dual to a magnetic
$k$-string and investigate whether this can compute the tensions and
world-sheet parameters reliably.

It is now well understood in a variety of different confining
backgrounds \cite{hk} 
that $k$-string tensions with $k$ of order $N_c$ in 
large $N_c$ theories, are computed by expanded brane
configurations. A collection of multiple probe F/D-strings 
can blow up into higher
dimensional D-branes by a version of the Myers effect, wrapping
topologically trivial cycles. A similar, very closely related
phenomenon also occurs for Wilson loops in general tensor
representations involving sources of varying $N$-ality in large $N_c$
gauge theories \cite{dfiol,malpol,yama}. In all these cases, the
expanded brane configuration carries a net $k$-string
charge by virtue of world-volume electric or magnetic fields.

\begin{figure}[h]
\begin{center}
\epsfig{file=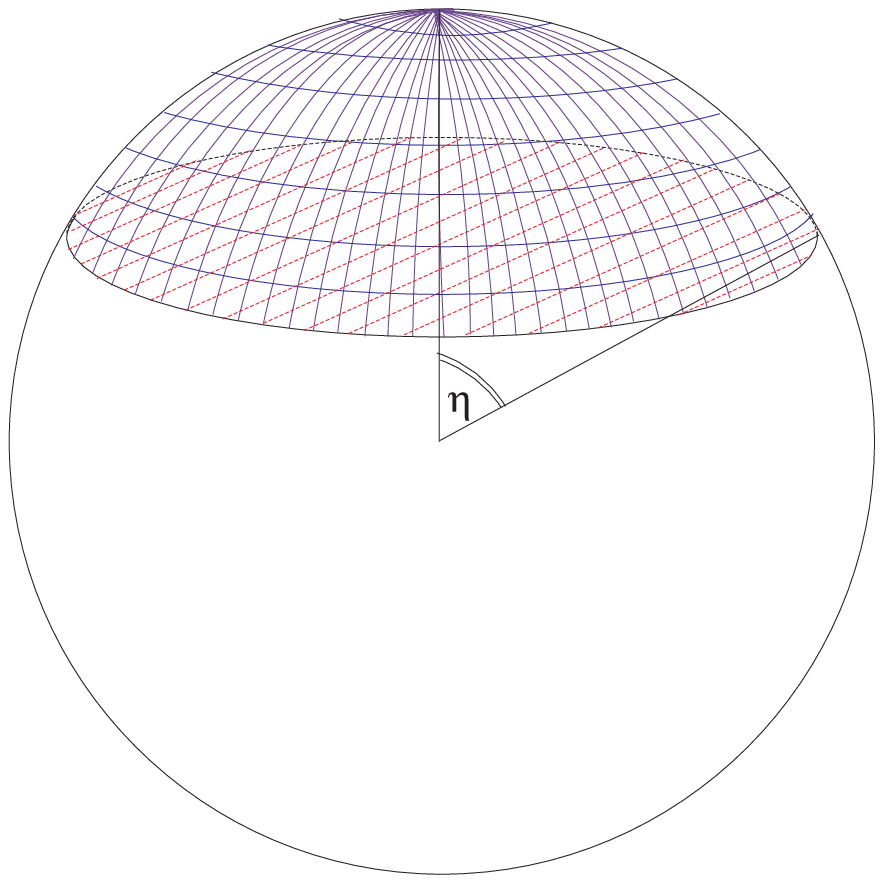, width = 5 cm}
\label{sferetta}
\end{center}
\noindent\small{\bf Figure 4:}  The minimal energy configuration for the
    probe D3 brane in the $\vec{w}$ space 
is given by the red disc and the blue ``polar cap'' shown in the
figure. The ``polar cap'' part is located very close to the D5 sphere.
\end{figure} 

In our large $N_c$ dual to the Higgs vacuum, we look for candidate branes that 
correspond to $k$-vortices with 
\beq
k\to\infty\,,N_c\to\infty\,\quad{\rm and}\,\quad{k\over N_c}\,\quad{\rm fixed}.
\eeq
The most natural object is a D3-brane with topology
$\mathbb{R}^{1,1} \times S^2$, and a  nonvanishing (magnetic) 
$F_2$ flux along
its compact directions located near the D5 shell. 
Our candidate probe D3-brane
is sketched in Figure 4. It is located at $\vec y=0$ 
and has the topology of an $S^2$ (but not the shape) in the $\vec w$
space. We believe this to be the correct configuration for large
enough $k$. For finite or small $k$, the $S^2$ of the probe D3-brane 
will be a small, smooth,
squashed sphere located at some polar angle along the D5-sphere. 
As $k$ is increased, the squashed sphere grows in size with
$k$. Most of this probe D3 sphere will want to stay near the D5 shell
where the $B_2$ field is concentrated, in order to minimize its
tension. The brane will, however, remain blown up due to the magnetic
field on its worldvolume which provides it with the requisite D-string
charge. The candidate probe brane also breaks the $SO(3)$ isometry to
$U(1)$ rotations around the $w_3$ axis, as we expect for the
$k$-vortices.

\subsubsection{A warm-up}
%Due to Myers effect \cite{myers}, a collection of probe $D1$ branes
%nearby the $D5$ shell 
%will blow up in a
%$D3$ with the topology of $\mathbb{R}^{1,1} \times S^2$ with some
%$F_2$ flux in the compact directions \cite{hk}. 
%We will see that in this case the blow up has just the topology of an $S^2$
%and not the shape. It corresponds to a configuration 
%at $\vec{y}=0$ and whose shape in the $\vec{w}$ space is
%shown in Fig.~\ref{sferetta}.
As a warm-up excercise, 
let us compute the DBI action for a spherically symmetric
configuration at constant $|\vec{w}|$, with $k$ units of uniform flux
on top of it. This will have higher tension than the D-brane in Figure 4. We
choose a constant world-volume magnetic field $F_2$ along the 
$(\theta_w, \phi_w)$ directions and proportional to the volume form of
$S^2$,
\beq
F_2 = {k\over 2}\sin\theta_w d\theta_w\wedge d\phi_w.
\eeq
The magnetic field 
induces a D-string charge $k$ for the spherical D3-brane, through
its Chern-Simons coupling
\beq
S_{\rm CS}= { 1\over (2\pi)^3\alpha^{'2}}
\int_{S^2 \times {\mathbb R}^{1,1}} 2\pi \alpha'F_2 \wedge C_2 =  {1\over
  2\pi\alpha'}\,k\,\int_{{\mathbb R}^{1,1}} C_2.
\eeq
The tension for the spherical D-brane will be obtained by minimizing
the DBI action with respect to $w$.
\beq 
S_{\rm DBI}=4 \pi \int d^2 x  \left(
\frac{Z_x^{-1/2} e^{-\Phi}}{(2 \pi)^3 \alpha'^2} \sqrt{Z_\Omega w^4 +4 \pi^2
\left( \frac{k \alpha'}{2} - \frac{N \alpha'}{2}
 \frac{1}{1+\frac{(w-r_0)^2}{\rho_c^2}}  \right)^2 
     } \right)
\, 
.
\eeq
%Near $w \approx r_0$, it reads:
%\beq  \int d^2 z  \left(
%\frac{2 r_0\sqrt{\rho_-^2+\rho_c^2}}{( \pi)^2 \alpha'^2 g R^2}
% \sqrt{ \frac{R^4 r_0^2 \rho_-^2}{4(\rho_c^2 + \rho_-^2)^2}
% +4 \pi^2 \left( \frac{k \alpha'}{2} - \frac{N \alpha'}{2}  \frac{\rho_c^2}{\rho_c^2 + \rho_-^2}  \right)^2 
%   } \right) \, .\eeq
The tension is minimized at $w\approx r_0$ (in the large $N_c$ limit)
and we obtain for the $SO(3)$ symmetric setup
\beq 
T_{\rm D3}\approx 2\pi \frac{m^2}{g^2_{\rm YM} } N_c (N_c-k) \, . \label{unstab} 
\eeq
We will see that the 
tension of the configuration in Figure 4 will be lower than the above
and so the $SO(3)$ symmetric $k$-string cannot be stable.

\subsubsection{The $k$-vortex}

Now let us compute the energy of the configuration in
Fig. 4. It consists of two parts: i) one which is a piece of
a sphere subtending the solid angle parametrized by
\beq
0\leq \theta_w\leq \bar \eta_k\,,\quad{\rm and}\quad 0\leq \phi_w\leq 2\pi,
\eeq
that we refer to as the polar cap, and ii) a disc glued to the
bottom of the cap. The two parts are distinguished by the fact that
polar cap lies close to the D5 shell at $w \approx r_0$ where the
$B_2$ field reaches its maximum. Since $B_2$ is non-zero only within a
thin region (\ref{flatb2}) of width $\rho_c\ll r_0$, the disc portion
of the expanded brane only sees a vanishing antisymmetric tensor
potential. In fact the geometry seen by the disc in the interior of
the D5-sphere is basically flat. As a consequence of this, the polar
cap can minimize its tension by having a magnetic field switched on
that completely cancels the pullback of $B_2$. 
This means that the polar cap is close to the D5-sphere and is
practically tensionless (suggesting that it is possibly
dissolved in the D5). In fact the entire tension of the configuration
arises from the disc. The disc itself cannot shrink since its boundary
must match on to the boundary of the polar cap, and the size of the
latter is fixed by the net D-string charge. Also, in the absence of
any $B_2$ field in the interior of the D5-sphere, the disc has no
world-volume magnetic field.

Let us first discuss the polar cap. The D3 world-volume 
is parameterized by the coordinates $(x_0,x_1,\phi_w,\theta_w)$.
With this portion of the brane at constant $|\vec{w}|\approx r_0$ and
$\vec y=0$,
we take the $F_2$ field to be proportional to the volume form of
the two-sphere
\beq 
F_2 =\frac{k}{ (\cos \bar{\eta}_k -1 )}  \sin \theta_w (d \theta_w
\wedge d \phi_w) \,\eeq 
where $0\leq \theta_w \leq \bar{\eta }_k$ and $0\leq \phi_w \leq 2
\pi$. The normalization is chosen so that this yields a D-string
charge $k$ for the blown up D3-brane. The D1-charge density is 
concentrated entirely in the polar cap portion.

The DBI action for the probe D3-brane reads
\beq S_{\rm cap}=\frac{1}{(2 \pi)^3 \alpha'^2} \int d^4 \xi \left\{ e^{-\Phi}
\sqrt{
 (- {\rm det} (G_{ab}+B_{ab}+2 \pi \alpha' F_{ab} ) ) } \right\}. \eeq
where
\begin{eqnarray}
G_{ab}+B_{ab}+2 \pi \alpha' F_{ab} &&=\\\nonumber 
&& \left(\begin{array}{cccc}
-Z_x^{-1/2} & 0& 0& 0\\
 0 & Z_x^{-1/2} & 0 & 0\\
 0  & 0 & Z_\Omega^{1/2}  w^2  & (2 \pi \alpha' F_{\theta \phi} +B_{\theta \phi}) \\
0  & 0 &  - (2 \pi \alpha' F_{\theta \phi} +B_{\theta \phi}) 
& Z_\Omega^{1/2} w^2 \sin^2 \theta_w  \\
\end{array}\right) \, .
\end{eqnarray}
From this we find the tension of the cap to be 
\begin{eqnarray}
&&S_{\rm cap}=\\\nonumber
&&\int d^2 x \, d \phi_w \,  d \theta_w\, \sin \theta_w \left(
\frac{Z_x^{-1/2} e^{-\Phi}}{(2 \pi)^3 \alpha'^2} 
 \sqrt{Z_\Omega w^4 +4 \pi^2  
\left( \frac{k \alpha'}{1-\cos \bar{\eta}_k} - 
 \frac{N \alpha'/2}{1+\frac{(w-r_0)^2}{\rho_c^2}}\right)^2  } \right)
\, .
\end{eqnarray}
The action for the polar cap 
needs to be extremized with respect to both $w$ and $\bar\eta_k$.
We find that there is a minimum at which the tension vanishes exactly
for 
$|\vec{w}|=r_0$ and 
\beq (1-\cos \bar{\eta}_k) = \frac{2 k}{N_c} \, 
. \label{etabar} \eeq
Since the action is positive definite and it vanishes exactly at the
above values of $w$ and $\bar\eta_k$, we have found the global minimum
of this contribution to the tension. The pullback of $B_2$ is exactly
cancelled by the magnetic field and the volume of the cycle goes to
zero near $w=r_0$ due to vanishing $Z_\Omega$ and the brane becomes
tensionless. At this point
we expect significant stringy corrections and the DBI approach is
invalid. These will likely change the tension for the polar cap, but
it is not clear whether the result will become comparable to the much
larger contribution to the tension from the disc at the bottom of the
polar  cap.

The disc lies for most of its extension at $|\vec{w}|-r_0\gg\rho_c$.
In this limit the relevant metric (at $\vec{y}=0$) is
\beq 
ds^2\big|_{w-r_0\gg \rho_c}= \frac{w^2-r_0^2}{R^2_{\rm AdS}} dx_\mu dx_\nu
\eta^{\mu \nu}+\frac{R^2_{\rm AdS}}{w^2-r_0^2} 
 (dw^2+w^2(d \theta_w^2+ \sin^2 \theta_w d \phi_w^2)) \, \eeq
and the dilaton is simply $e^\Phi=g_s$. The $B_2$ field is also small
in this region.  Hence,
the disc portion of the probe D3 has two space-time directions $(x_0,x_1)$,
and two directions in the $\vec{w}$ space, without fluxes. The warp
factors from the two different subspaces cancel out and then 
the resulting DBI action is equivalent to
the one for a membrane in flat space with fixed perimeter.
So the tension of the disc is given by its area in flat space,
\beq 
T_{\rm D3}\approx 
\frac{1}{(2 \pi)^3 \alpha'^2} \frac{1}{g_s}\, \left (\pi r_0^2 \sin^2
  \bar{\eta_k}\right) \,
=\,\frac{m^2 }{2 g_s} k(N_c-k) \, . \label{d3t}\eeq 
which is indeed less than the tension of the $SO(3)$ symmetric ansatz
in Eq.~(\ref{unstab}). This gives the tension of the $k$-vortex and
remarkably, matches our weak coupling semiclassical results and the
softly broken ${\cal N}=2^*$ formula.
In this analysis we neglected the small region at $w-r_0 \lesssim\rho_c$;
a more careful DBI analysis yields
\beq 
T_{\rm D3}=\frac{1}{g_s (2 \pi)^2 \alpha'^2} \int_0^{r_0 \sin \bar{\eta}_k} 
\sqrt{\frac{1}{(r_0 \cos \bar{\eta}_k)^2+s^2} \left( s^2+ (r_0 \cos \bar{\eta}_k)^2 \frac{(w-r_0)^2}{ (w-r_0)^2 +\rho_c^2} \right)} \,  s
ds  \, .\eeq 
The formula is obtained following a non-trivial cancellation between
the dilaton and the metric warp factors. Clearly for $|w-r_0|\gg
\rho_c$, the result is given by the area of the flat disc. It is a
good approximation to ignore $\rho_c$ relative to the disc radius,
since 
\beq
{\rho_c\over r_0} =\sqrt{g_s\over N_c \pi} \ll 1.
\eeq
It is straightforward to check
 that a more careful extremization 
does not change the result at the leading order in $g_s$.

A rather interesting cross-check of the picture above results when 
 one determines the tension of  the same  kind of configuration 
(a polar cap with a disc glued) with generic values of $\bar{\eta}_k$,
 i.e. where $\bar \eta_k$ is allowed to be a free parameter instead of
 being determined by the $k$-string charge. 
The resulting tension formula is then
\beq 
T_{\rm D3}(\bar\eta_k, k)=
\frac{m^2}{2 g_s} \left( \frac{N^2_c \sin^2 \bar{\eta}_k }{4} 
+ N_c \left| N_c \frac{1-\cos \bar{\eta}_k}{2} - k \right| \, 
\right) \, ,\eeq
which, indeed always has a minimum at the value of 
$\bar{\eta}_k$ given by Eq. (\ref{etabar}),
as shown in Figure 5.

\begin{figure}[h]
\begin{center}
\epsfig{file=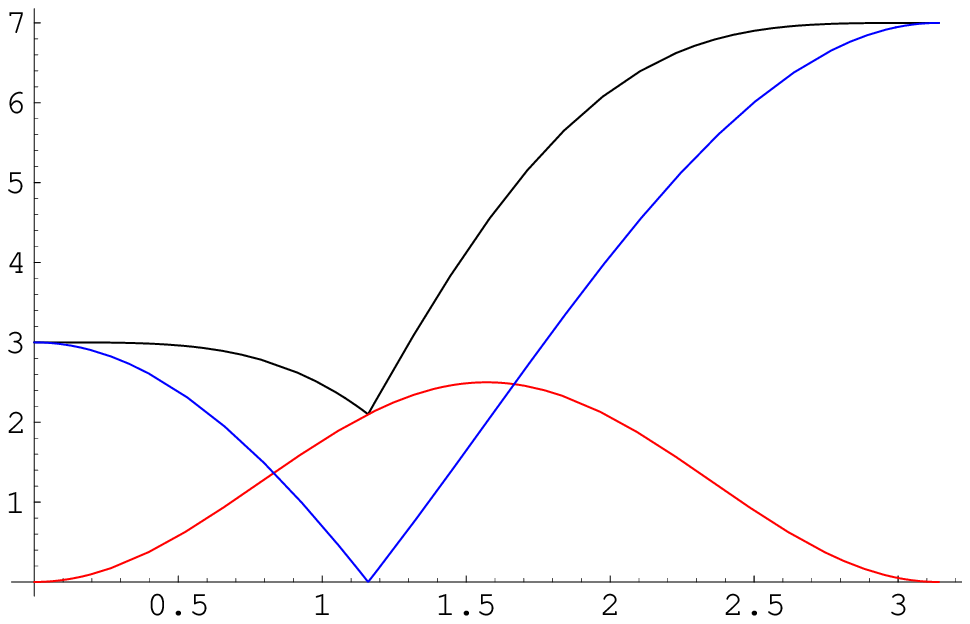, width =2.5in}
\end{center}
\noindent\small{\bf Figure 5:} Energy as a function of $\bar{\eta}_k$
for $N=10$, $k=3$. The blue line is proportional to the energy of the
``polar cap'', 
the red one to the energy of the flat disc and the black line is the
sum of the two contributions. 
The minimum is given by Eq. (\ref{etabar}).
\end{figure}

\subsubsection{Theta term in world-sheet sigma model}

Since our expanded D3-brane configuration breaks the $SO(3)$ isometry
to $U(1)$, it follows that small fluctuations of the orientations of
the configuration along the D5-sphere, will lead to a $1+1$
dimensional sigma model with target space ${\mathbb {CP}}^1$. As
usual, all potentially interesting physics 
lies in the theta angle of this sigma model.
From this picture it is straightforward to find also the
   $\theta_{1+1}$ of the effective $S^2$ sigma model.
This comes from the following Chern-Simons coupling in the D3-brane theory,
\beq 
S^{\rm CS}_{\rm cap}= \frac{1}{(2 \pi)^3 \alpha'^2} \int_{\rm cap} C_2 \wedge
( 2 \pi\alpha' F_2) \, . 
\label{wezzuzzu}\eeq
For $\theta_{3+1}\neq 0$, the components of $C_2$ tangential to the
polar cap, can be read off from (\ref{flatb2}) and (\ref{thetarotaz}).
Note that the only non-zero contribution comes from the polar cap, since
it has a magnetic field,
\beq 
F_2=\frac{N_c}{2} \sin \theta_w (d \theta_w \wedge d \phi_w) \, . 
\eeq
Let us denote with $\vec{n}_w$, the position of the North Pole
at the center of the polar cap  and for simplicity, 
let us orient $\vec{n}_w$ 
 in the $w^3$ direction. We also denote the 
position vector of any point on the polar cap as
\beq 
\vec{p}= (\sin \theta_w \cos \phi_w, \sin \theta_w \sin \phi_w, \cos
\theta_w) \,.
\eeq 
Now, we want to consider the effect of an infinitesimal displacement
of the entire polar cap following from the action of a rotation
generator in $SO(3)/U(1)$. We allow this displacement to have an
adiabatic dependence on the non-compact coordinates $x^0$ and $x^1$.
Under the infinitesimal change $\vec n_w$ transforms as
\beq 
\vec{n}_w \rightarrow \vec{n}_w+{\partial_0 \vec{n}_w}\, dx^0+
{\partial_1 \vec{n}_w}\, dx^1 \, . \eeq
%where the derivatives of $\vec{n}_w$ are in the  $\hat{x},\hat{y}$
%directions of the $\vec{w}$ space. 
The corresponding displacement for a generic point $\vec{p}$ on the
polar cap is, 
\beq 
\vec{p} \rightarrow \vec{p}+
{\partial_0 \vec{p}}\,dx^0+
{\partial_1\vec{p}} \,dx^1\,.
\eeq
It is easily seen that the variations in the position vector $\vec p$
are related to the change in $\vec n_w$ as
\beq 
{\partial_s \vec{p}}=\left(\vec{n}_w \times {\partial_s \vec{n}_w} 
\right) \times \vec{p} \, .  \eeq
Armed with these relations between the variations in $\vec p$ and those in
$\vec n_w$, we turn to the Chern-Simons term (\ref{wezzuzzu}) in the
D-brane action. First of all the pullback of $C_2$ can be
re-expressed in terms of the unit vectors corresponding to the
location of each point on the polar cap
\beq 
P[C_2]= - \frac{  \theta_{3+1} \alpha' N_c }{2 }\,   
\left( {\partial_0 \vec{p}}  \times {\partial_1\vec{p}}
\right)  \cdot  \vec{p}   \,\,  d x^0 \wedge d x^1 \, ,
\eeq
which can be written in terms of the unit polar vector $\vec n_w$ as
\beq 
P[C_2]=- \frac{  \theta_{3+1} \alpha' N_c }{2 }  \cos \theta_w 
%(1+ \cos 2 \phi_w  \sin^2 \theta_w) 
\left({\partial_0 \vec{n}_w} \times {\partial_1 \vec{n}_w}
\right)  \cdot  \vec{n}_w    \,  d x^0 \wedge d x^1 \, .
\eeq
We substitute this in Eq. (\ref{wezzuzzu}) and integrate over the
polar angles of the cap to obtain the topological term in the
world-sheet sigma model of the $k$-vortex 
 \beq 
{\cal L}_\theta = -\frac{k(N_c-k) \, \theta_{3+1} }{8 \pi}\,
 \epsilon^{s r} \vec n_w \cdot \left(\partial_s \vec n_w \times  \partial_{r}
   \vec n_w \right).
 \eeq
Once again, despite potential issues with regard to the high curvatures
in the vicinity of the D5-sphere, the answer is in agreement with
the physics at $g^2_{\rm YM}\ll 1$ and is invariant under shifts of
the Yang-Mills theta angle by multiples of $2\pi$. As before, one
likely reason for the robust nature of the result is that it
originates from the Chern-Simons couplings of the D3-brane.

The coefficient of the kinetic term $\tilde{B}_{N_c,k}$ is a trickier issue in the probe D3
approach. We have already remarked that this will flow in the
quantized sigma model. Nevertheless it is an object that we can
formally estimate.
The contribution to this quantity from the cap is zero;
there is a non-zero contribution from the disc that we will estimate here.
Let us parameterize the disc in the $\vec{w}$ space with the coordinate $\vec{q}$: 
\beq \vec{q}= (s \cos \phi , s  \sin \phi, r_0 \cos \bar{\eta}_k) \, .\eeq
The infinitesimal variation of $\vec{q}$ is
\beq 
\frac{\partial \vec{q}}{\partial x^s}=
\left(\vec{n}_w \times \frac{\partial \vec{n}_w}{\partial x^s} \right) \times \vec{q} \, ,  \eeq
which upon inserting in the DBI action, yields
\beq 
{\cal L}_{\rm kin}=\tilde{B}_{N_c,k} \left( \frac{\partial \vec{n}_w}{\partial
    x^s}  \right)^2 \, ,
\eeq
where
{ \small \beq \tilde{B}_{N_c,k}=\int d s d \phi \left(
\frac{R^4_{\rm AdS} 
s \left(s^2 \cos ^2(\phi )+\cos ^2(\bar\eta_k) r_0^2\right)
\sqrt{\frac{\left(\left(r_0-w\right)^2+\rho _c^2\right) 
   s^2+\cos ^2(\bar\eta_k ) r_0^2 \left(r_0-w\right)^2}{\left(s^2+\cos
     ^2(\bar\eta_k ) r_0^2\right) 
\left(\left(r_0-w\right)^2+\rho
   _c^2\right)}}}{16 g_s 
\pi ^3 \alpha'^2 \left(w+r_0\right)^2 \left(\left(r_0-w\right)^2+\rho
  _c^2\right)} \right) \, , 
\eeq }
which scales as $1/\sqrt{g_s}$ and not as $1/g_s$ as we would expect
from the field theory and also from the D1 probe calculation.
The contribution from the interior of the disc is $\mathcal{O}(g_s^0)$.
The leading contribution of order $\mathcal{O}(1/\sqrt{g_s})$ comes from 
the boundary of the disc at $w-r_0 \approx \rho_c$.
This is the region nearby the intersection between the cap and the
disc and indeed there we can not trust our guess for the shape of the  D3  
brane probe.

\subsection{Relation to the Baryon Vertex}

In the context of gravity duals of gauge theories, there exists a
close relationship between baryons and flux tubes. Flux tubes are made
of the same material as baryons, first seen in the context of the
baryon vertex in ${\cal N}=4$ theory \cite{bion, wittenbaryon} and
subsequently for gravity duals of confining gauge theories
\cite{calguij1,calguij2,sean}. In all these examples, a baryon vertex
with $N_c$ strings attached, represented by a wrapped D5-brane for
example, can be deformed and pulled apart into groups of constituent quarks
connected by a flux tube. The portion of the D5-brane that looks like
a flux tube in the gauge theory is obtained by the 5-brane wrapping an
$S^4 \subset S^5$. 

Polchinski and Strassler \cite{ps} argued that a D5-brane baryon
vertex of the UV ${\cal N}=4$ theory, when taken to the IR by moving
towards the interior of the ${\cal N}=1^*$ geometry, eventually
meets the dielectric 5-brane spheres in the interior. In the confining
vacuum when the baryon vertex is moved past the NS5-sphere, by the
Hanany-Witten process of brane creation \cite{hwitten}, the D5-brane
baryon vertex   
turns into a D3-brane ball filling the  space inside the
NS5-sphere. Following a similar logic applied to a D5-brane wrapping an
$S^4$ inside the $S^5$ in the far UV geometry, in the IR
we would expect the ${\mathbb Z}_{N_c}$ 
flux tube to be a D3-brane with
world-volume ${\mathbb R}^{1,1}\times {\cal D}_2$ where the ${\cal D}_2$
is a disc stretching inside and ending on the dielectric sphere. The
magnetic flux tube would essentially be the same type of object,
obtained by S-duality on the confining vacuum. It is encouraging to
see that the crucial portion of our candidate $k$-vortex, 
the expanded D3-brane,
is precisely such a D3-brane disc. The tension of the magnetic flux tube
arises entirely from this disc. 
Nevertheless, the polar cap was crucial for providing the
boundary condition that stabilized the disc, and for providing the
magnetic field responsible for $k$ units of D-string
charge. It would be interesting to understand better, the precise connection
between the two slightly differing pictures.

\subsection{Confining Vacuum}

We conclude our discussion on flux tubes in the Polchinski-Strassler
dual, with a brief analysis of $k$-strings in the
confining vacuum. First of all we note that 
the tension of the $k$-string in the confining phase
at strong coupling $g_{\rm YM}^2 N_c \gg 1$ is simply the S-dual of the
magnetic $k$-string tension in the Higgs vacuum at weak coupling or
$N_c/g_{\rm YM}^2 \gg 1$. Hence we learn that at least at large $N_c$,
and large 't Hooft coupling, the confining $k$-strings must
obey a Casimir scaling law,
\beq
T_{N_c,k}^{\rm confining}= m^2 \,{g^2_{\rm YM}\over 8\pi}\,k(N_c-k).
\eeq
A direct confirmation of this from the corresponding expanded D3-brane
in the confining vacuum geometry would be useful, but we leave this
for future study. The Casimir scaling for confining string
tensions is in contrast to previously encountered 
sine laws and approximate sine
laws in other confining theories \cite{ds, hsz, hk}. 
The confining vacuum is manifestly $SO(3)$ invariant in the absence of
any VEVs for the adjoint scalars. 

Below we outline the calculation of the 
tension for a $k=1$ flux tube in the confining vacuum (first done in
\cite{ps}) to see how it is consistent with the action of S-duality.
The configuration corresponding to the confining vacuum is an
NS5-brane 
wrapped on a sphere. The corresponding supergravity background
can be found via S-duality from the Higgs vacuum  with 
the fields and the parameters transforming as
\begin{eqnarray}
&&g_s \rightarrow \tilde{g}_s=\frac{1}{g_s}, \qquad 
\alpha' \rightarrow \tilde{\alpha}'=g_s \alpha', \qquad
\exp (\Phi) \rightarrow \exp (\tilde{\Phi}) = \exp (-\Phi),\\\nonumber 
&&ds^2 \rightarrow d \tilde{s}^2=g_s \exp (- \Phi) ds^2,\qquad
(B_2 , C_2) \rightarrow (\tilde{B}_2 , \tilde{C}_2) = (-C_2,B_2 )\, .
\end{eqnarray}
The profile functions for the new background metric are then
\beq 
Z_x=Z_\Omega=\frac{R^4_{\rm AdS}}{\rho_+^2 (\rho_-^2 + \rho_c^2)}, 
 \,\,\, Z_y=\frac{R^4_{\rm AdS}  (\rho_-^2 + \rho_c^2)}{\rho_+^2 \rho_-^4},\eeq
with the metric having the same form as (\ref{hmetric}). The 
parameters in the metric and the dilaton are,
\beq r_0=({\alpha}' m) \pi  {g}_s N_c, \qquad
 \rho_c=({\alpha}' m) \sqrt{\pi {g}_s N_c},\qquad
e^{2 \Phi}= {g}^2_s \frac{\rho_-^2+\rho_c^2}{\rho_-^2}.
\eeq
This is the background for $\theta_{3+1}=0$. The confining string of
the gauge theory is identified with the F-string (dissolved in the NS5
sphere), which couples directly to the string metric.
The action for the probe fundamental string is
\beq 
S_{\rm F1}=\frac{1}{2 \pi \alpha'} \int d^2 x \left\{ 
\sqrt{- {\rm det} (G_{ab}) }  + B_{ab} \right\}, \eeq
with the string oriented in the $x_0,x_1$ directions. Upon evaluating
this action we find 
\begin{eqnarray}
&&S_{\rm F1}=\frac{1}{2 \pi \alpha'} \int d^2 x
(Z_x^{-1/2} )\\\nonumber
&&= \frac{1}{2 \pi \alpha'} \int d^2 x
\frac{\sqrt{y^2+(w+r_0)^2}  \sqrt{y^2+(w-r_0)^2+\rho_c^2 }  }{
  R^2_{\rm AdS}}. 
\end{eqnarray}
The location of the minimum is the same as in the D5 case and the
resulting string tension 
\beq 
T_{\rm F1}=\frac{m^2 g_s N_c}{2},
\eeq
is exactly the S-dual of the vortex tension. At the radial position of the
string $|w-r_0|\approx \rho_c^2/2r_0 = \alpha' m/2 \ll r_0$, the radius of
the sphere with constant $w$ is $\sqrt{\alpha' \pi g_s N_c}$ which is
large in string units. It is interesting that the flux tube in the
strongly coupled confining vacuum appears to break the global $SO(3)$
invariance, as it is point-like on the sphere. This is
counter-intuitive, since the $SO(3)$ is an exact global symmetry of the
confining vacuum and not a colour-flavour locked transformation as in
the Higgs vacuum. So we do not expect the confining strings to have any
orientational zero modes. This should become manifest upon quantizing
the associated sigma model, whereby the quantum wavefunction spreads
over the entire classical moduli space and the classical zero modes are removed.
To obtain the classical sigma model for the flux tube, we can allow
the string to 
fluctuate in the directions tangential to the sphere and these would
give the action for the  ``classical orientational zero modes'' of the
confining string flux 
tube. From the Nambu-Goto action for the string we get 
\beq 
S_{\rm F1}=\int d^2 x \frac{1}{2 \pi \alpha'}
\sqrt{ {\rm Det} \left(\begin{array}{cc}
-Z_x^{-1/2}+(\partial_0 \vec{n}_w)^2 w^2 Z_{\Omega}^{1/2}  &
(\partial_0 \vec{n}_w) (\partial_1 \vec{n}_w) w^2 Z_{\Omega}^{1/2}  \\ 
 (\partial_0 \vec{n}_w) (\partial_1 \vec{n}_w) w^2 Z_{\Omega}^{1/2}
 & Z_x^{-1/2}+(\partial_1 \vec{n}_w)^2 w^2 Z_{\Omega}^{1/2}.  \\ 
\end{array}\right) }
\eeq
The effective action at the two-derivative level is 
\beq S_{\rm F1}=  \int d^2 x
\left( \frac{g_s N_c}{4} (\partial_s \vec{n}_w)^2  \right).  \eeq
It is nice to see that the classical coupling is exactly S-dual to the
one for the D-string. The theta dependence of the sigma model action 
could provide us with
further clues about the worldsheet dynamics of these flux
tubes. However, the confining background above is for
$\theta_{3+1}=0$. Shifting the Yang-Mills theta angle by multiples of
$2\pi$ changes the vacuum to one in an oblique-confining phase and the
NS5 brane to a $(1, n)$ 5-brane. If we dial $\theta_{3+1}$ 
as was done in the
Higgs vacuum using an $SL(2,{\mathbb R})$ shift, this does not alter
the $B_2$ field, although it does change the R-R potential
$C_2$. However, the former does not have components along the NS5 sphere,
whilst only the
RR two form does and we do not know how this couples to the
F1 world-sheet in a simple way.

\section{Summary and further questions} \label{summary}

In this paper we have first studied  solitonic $k$-vortices
in the Higgs vacuum of the $SU(N_c)$ $\mathcal{N}=1^*$ theory with
equal adjoint masses, transforming under an $SO(3)_{C+F}$ symmetry group. 
We have found that for every $k$ and $N_c$ the vortex
world-sheet theory is a non-supersymmetric $S^2$ sigma model. Perhaps the most
interesting feature of the two dimensional world-sheet theory is the 
relation between its theta angle and the four dimensional one,
$\theta_{1+1}= k(N_c-k)\theta_{3+1}$. This has very specific
implications for the IR dynamics of the sigma model which is
asymptotically free and for general values of $\theta_{1+1}$, has a
mass gap with the spectrum consisting of a triplet of $SO(3)$. 
When $\theta_{1+1}=\pi$ however, the theory is integrable and the 
spectrum consists of massless doublets of $SO(3)$ and the theory flows
to a $c=1$ conformal fixed point. The doublets of $SO(3)$ which are
confined into meson-like triplet states become deconfined and massless
at $\theta_{1+1}=\pi$. However this value of the world-sheet theta
angle corresponds to a seemingly non-special value of the spacetime
theta angle. We have speculated on the possibility that the 
values: 
\[ \theta_{3+1}=\frac{\pi}{k(N_c-k)}  \, ,\] 
may correspond to a level crossing
in the semiclassical spectrum of massive, 
mutually non-local monopole-dyon states of the parent 
${\cal N}=2^*$ theory in the Coulomb phase. 
These states would be confined in
the Higgs vacuum and appear as $SO(3)$ doublets bound to a
$k$-vortex. We have not presented any evidence for this, but clearly,
further study of the relation between the vortex world-sheet spectrum
and the spectrum of the four dimensional theory will reveal
interesting physics.

The relation between the theta angles on the worldsheet and spacetime
also implies that instantons of  charge one
in the $k$-string $\mathbb{CP}^1$ sigma model correspond to
multi-instantons in the four dimensional theory.

Yet another feature of our non-Abelian string solutions is that their
tensions, evaluated numerically, are extremely well approximated by
the Casimir scaling law, 
\[ T_{N_c,k} = \frac{2\pi m^2 k (N_c-k)}{g^2_{\rm  YM}} \, , \] 
for large $N_c$. This is remarkable in that the Casimir scaling
formula is known to be valid for vortices in the softly broken ${\cal
  N}=2^*$ theory because of ``almost ${\cal N}=2$ SUSY'' and the
resulting string solutions are BPS. Our solutions are far from BPS and
the agreement with the BPS tension formula, at large $N_c$ is
surprising and needs further explanation. 
The semiclassical Higgs vacuum is mapped by
S-duality to the confining vacuum at strong 't Hooft coupling and so
confining string tensions in this regime will also obey Casimir scaling.

We have identified the supergravity duals of the vortex strings 
in the large $N_c$ limit of ${\cal N}=1^*$ theory. The dual IIB 
string background,
due to Polchinski and Strassler, has a parametric regime of validity 
$N_c/g_s\gg 1$, which includes the semiclassical regime of weak gauge
coupling. For this reason, and more explicitly from the form of the
metric itself which is sourced by a D5-brane in the IR, 
it is expected that the IR physics of the Higgs vacuum lies outside
the regime of supergravity due to large curvatures. We find it
surprising therefore, that we were not only able to identify the
candidate objects dual to $k$-vortices in the large $N_c$ limit, as
expanded D3-branes, but also able to compute their tensions and find
an exact Casimir scaling in agreement with the semiclassical results. 
As a bonus we were able to reproduce the semiclassical
relation between the Yang-Mills theta angle and the worldsheet theta
angle from the candidate wrapped, D-brane configurations. It would
definitely be useful to understand better the reason for this
agreement and connect to some kind of large $N_c$ BPS property.

The Higgs vacuum of ${\cal N}=1^*$ theory, in the large $N_c$ limit, 
has been argued to provide a deconstruction of six dimensional
supersymmetric gauge theory compactified on a fuzzy sphere
\cite{da}. In this picture, the vortices may be reinterpreted as 
noncommutative instantons \cite{Nekrasov} of the six-dimensional theory. This
presents a potentially fruitful arena for systematically investigating
the Higgs vacuum vortices, at least in the large $N_c$ limit \cite{inprog}
, and may explain some of the surprising features above at large
$N_c$. Finally, a closely related situation arises in the beta
deformation of ${\cal N}=4$ theory at special values of $\beta$,
wherein the resulting theory in its Higgs phase has been argued to
deconstruct Little String Theory \cite{dorey}. 
\\\\
{\bf Acknowledgements:} The authors would like to thank A. Armoni, C. Hoyos-Badajoz, K. Konishi, 
B. Lucini, C. Nu$\tilde{\rm n}$ez and A. Yung for discussions.

\startappendix
\Appendix{``Near Shell'' Background: Flat D5 and NS5 with D3 charge}
Below we quote the form of the ``near-shell'' background in the Higgs (D5)
and confining (NS5) solutions of Polchinski and Strassler. In the
Higgs (confining) phase, the
``near-shell'' solutions are directly read off from the 
metric in string frame for  $p$ flat D5-branes (NS5 branes) with 
D3-brane charge bound to them \cite{AOSJ,BMM}:
\\\\
{\bf D5/D3 background:}
\beq 
ds^2=\frac{dx_\mu \, dx_\nu \eta^{\mu \nu}}{\sqrt{1+\frac{S^2}{\alpha'^2 u^2}}}
+\frac{\sqrt{1+\frac{S^2}{\alpha'^2 u^2}}}{\frac{S^2}{\alpha'^2 u^2}
  \cos^2 \varphi+1}(d x_4^2+d x_5^2)+
\alpha'^2 \sqrt{1+\frac{S^2}{\alpha'^2 u^2}} (du^2+u^2 d \Omega_3^2) \, \eeq
where 
\beq S=\sqrt{ \frac{g_s p \alpha'}{ \cos \varphi} } \, ,\eeq
and $\tan \varphi$ is proportional to the density of D3 brane charge
dissolved on the D5-branes. 
The dilaton and the $B$ fields are given by:
\beq e^{2 \Phi}=g^2_s \frac{\alpha'^2 u^2 }{S^2 \cos^2 \varphi+
  \alpha'^2 u^2} \, , \qquad 
 B_{45}=-\frac{S^2 \sin \varphi \cos \varphi}{S^2 \cos^2 \varphi+
   \alpha'^2 u^2} \, .\eeq 
The RR potentials are:
\begin{eqnarray}
&&C_2=\pm 2 \frac{ S^2 \cos \varphi}{g_s} \sin^2 \theta \cos \phi_1
\, d \theta \wedge d \phi_2 \, ,  \\
&&C_4=\mp \frac{2 S^2 \sin \varphi}{g_s} \, \frac{ r^2 +S^2/2 \cos^2
  \varphi}{ r^2 +S^2 \cos^2 \varphi} 
\sin^2 \theta \cos \phi_1 d x_5 \wedge d x_4 \wedge d \theta \wedge d
\phi_2 
\\\nonumber
&&\pm \frac{\sin \varphi}{g_s} \, \frac{r^2}{r^2+S^2} dx^0 \wedge d x^1
\wedge dx^2 \wedge dx^3 \, ,
\end{eqnarray} 
where $(\theta,\phi_1,\phi_2)$ are the standard coordinates in $S^3$.

In the decoupling limit $\alpha' \rightarrow 0$, $\tan \varphi
\rightarrow \infty $ with $\alpha' \, \tan \varphi=b$ held
constant, after the rescaling $\tilde{x}_{4,5}=\tan \varphi x_{4,5}$,
the solution reads: 
\begin{eqnarray}
&&ds^2=\alpha' \left( \frac{u}{g_s p a} \left(
    \frac{d\tilde{x}_4^2+d\tilde{x}_5^2} 
{1+ a^2 u^2}+ dx_\mu \, dx_\nu \eta^{\mu \nu}\right)+ \frac{g_s p a}{u}
(du^2+u^2 d \Omega_3^2) \right)\, , 
\label{flatd5limit} \\\nonumber
&&e^{2 \Phi}=g^2_s \frac{a^2 u^2}{1+a^2 u^2}\, , \qquad
 B_{45}=-\frac{\alpha'}{g_s p a^2} \frac{1}{1+a^2 u^2} \, ,
\end{eqnarray} 
where we have defined  
\beq a=\sqrt{\frac{\alpha' \tan \varphi }{g_s p }} \, .
\eeq

The identification used in the Polchinski-Strassler 
solution to interpolate between the near-shell and the asymptotic metric is,
\beq 
u=\frac{\rho_-}{\alpha'} , 
\qquad a= \frac{1 }{p \, m \sqrt{g_s N_c \pi} } \, \qquad
\tilde{x}^{4,5}=\frac{1}{ p \,   \alpha' m^2 \pi N_c} w^{1,2} \, .
\eeq 
\\\\
{\bf NS5/D3 background:}
The metric in string frame for  $p$ flat NS5-branes with D3-brane charge 
(in the appropriate decoupling limit) is,
\beq 
ds^2=\frac{{\alpha}' {g_s}^2}{p a^2}  \left( \frac{d
    \tilde{x}_4^2 + d \tilde{x}_5^2}{\sqrt{1+a^2 u^2}} + 
\sqrt{1+a^2 u^2} d x_\mu dx_\nu \eta^{\mu \nu}\right)
+{\alpha}' p \frac{\sqrt{1+a^2 u^2}}{u^2} (du^2+u^2 d \Omega_3^2
) \, .
\eeq
The dilaton and the $C_2$ fields are:
\beq 
e^{2 \tilde{\Phi}}= {g_s}^2 \frac{1+a^2 u^2}{a^2 u^2 } , 
\qquad
C_2=-\frac{{g_s}^2 {\alpha}' }{a^2 p } \frac{1}{1+a^2 u^2} d
\tilde{x}_4 \wedge d \tilde{x}_5 \, .
\eeq 
In the confining vacuum, 
the identification used in the PS solution between 
near-shell and the asymptotic metric is,
\beq u=\frac{\rho_-}{{\alpha}' {g_s}} , \qquad a=
\frac{\sqrt{{g_s}} }{p \, m \sqrt{ N_c \pi} } \, \qquad 
\tilde{x}^{4,5}=\frac{1}{ p \,   {g_s} 
{\alpha}' m^2 \pi N_c} w^{1,2} \, .\eeq


\begin{thebibliography}{99}


%\cite{Seiberg:1994rs}
\bibitem{sw}
  N.~Seiberg and E.~Witten,
  %``Electric - magnetic duality, monopole condensation, and confinement in N=2
  %supersymmetric Yang-Mills theory,''
  Nucl.\ Phys.\  B {\bf 426}, 19 (1994)
  [Erratum-ibid.\  B {\bf 430}, 485 (1994)]
  [arXiv:hep-th/9407087].
  %%CITATION = NUPHA,B426,19;%%;

  N.~Seiberg and E.~Witten,
  %``Monopoles, duality and chiral symmetry breaking in N=2 supersymmetric
  %QCD,''
  Nucl.\ Phys.\  B {\bf 431}, 484 (1994)
  [arXiv:hep-th/9408099].
  %%CITATION = NUPHA,B431,484;%%


 

\bibitem{ano}
A. Abrikosov,
Sov.\ Phys.\ JETP {\bf 32}, 1442 (19yy).
%%CITATION = SPHJA,32,1442;%%

\bibitem{nielsen}
  H.~B.~Nielsen and P.~Olesen,
  %``VORTEX-LINE MODELS FOR DUAL STRINGS,''
  Nucl.\ Phys.\  B {\bf 61}, 45 (1973).
  %%CITATION = NUPHA,B61,45;%%



%\cite{Vafa:1994tf}
\bibitem{Vafa:1994tf}
  C.~Vafa and E.~Witten,
  %``A Strong coupling test of S duality,''
  Nucl.\ Phys.\  B {\bf 431}, 3 (1994)
  [arXiv:hep-th/9408074].
  %%CITATION = NUPHA,B431,3;%%




%\cite{Donagi:1995cf}
\bibitem{Donagi:1995cf}
  R.~Donagi and E.~Witten,
  %``Supersymmetric Yang-Mills Theory And Integrable Systems,''
  Nucl.\ Phys.\  B {\bf 460}, 299 (1996)
  [arXiv:hep-th/9510101].
  %%CITATION = NUPHA,B460,299;%%

\bibitem{Dorey:1999sj}
  N.~Dorey,
  %``An elliptic superpotential for softly broken N = 4 supersymmetric
  %Yang-Mills theory,''
  JHEP {\bf 9907} (1999) 021
  [arXiv:hep-th/9906011].
  %%CITATION = JHEPA,9907,021;%%


\bibitem{Dorey:2000fc}
  N.~Dorey and S.~P.~Kumar,
  %``Softly-broken N = 4 supersymmetry in the large-N limit,''
  JHEP {\bf 0002} (2000) 006
  [arXiv:hep-th/0001103].
  %%CITATION = JHEPA,0002,006;%%


\bibitem{ps}
  J.~Polchinski and M.~J.~Strassler,
  ``The string dual of a confining four-dimensional gauge theory,''
  arXiv:hep-th/0003136.

\bibitem{strassler}
M.~J.~Strassler,
  ``Messages for QCD from the superworld,''
  Prog.\ Theor.\ Phys.\ Suppl.\  {\bf 131}, 439 (1998)
  [arXiv:hep-lat/9803009].
  %%CITATION = PTPSA,131,439;%%
 
M.~J.~Strassler,
  ``Millennial messages for QCD from the superworld and from the string,''
  arXiv:hep-th/0309140.
  %%CITATION = HEP-TH/0309140;%%; 
  


%\cite{Montonen:1977sn}
\bibitem{Montonen:1977sn}
  C.~Montonen and D.~I.~Olive,
  %``Magnetic Monopoles As Gauge Particles?,''
  Phys.\ Lett.\  B {\bf 72}, 117 (1977).
  %%CITATION = PHLTA,B72,117;%%



\bibitem{HT}
  A.~Hanany and D.~Tong,
  %``Vortices, instantons and branes,''
  JHEP {\bf 0307} (2003) 037
  [arXiv:hep-th/0306150].
  %%CITATION = JHEPA,0307,037;%%

\bibitem{ABEKY}
  R.~Auzzi, S.~Bolognesi, J.~Evslin, K.~Konishi and A.~Yung,
  %``Nonabelian superconductors: Vortices and confinement in N = 2 SQCD,''
  Nucl.\ Phys.\  B {\bf 673} (2003) 187
  [arXiv:hep-th/0307287].
  %%CITATION = NUPHA,B673,187;%%

%\cite{Tong:2003pz}
\bibitem{dt}
  D.~Tong,
  %``Monopoles in the Higgs phase,''
  Phys.\ Rev.\  D {\bf 69}, 065003 (2004)
  [arXiv:hep-th/0307302].
  %%CITATION = PHRVA,D69,065003;%%

%\cite{Shifman:2004dr}
\bibitem{Shifman:2004dr}
  M.~Shifman and A.~Yung,
  %``Non-Abelian string junctions as confined monopoles,''
  Phys.\ Rev.\  D {\bf 70}, 045004 (2004)
  [arXiv:hep-th/0403149].
  %%CITATION = PHRVA,D70,045004;%%

\bibitem{HT2}
  A.~Hanany and D.~Tong,
  %``Vortex strings and four-dimensional gauge dynamics,''
  JHEP {\bf 0404}, 066 (2004)
  [arXiv:hep-th/0403158].
  %%CITATION = JHEPA,0404,066;%%



\bibitem{SYrev}
  M.~Shifman and A.~Yung,
  %``Supersymmetric Solitons and How They Help Us Understand Non-Abelian   Gauge
  %Theories,''
  Rev.\ Mod.\ Phys.\  {\bf 79} (2007) 1139
  [arXiv:hep-th/0703267].
  %%CITATION = RMPHA,79,1139;%%



\bibitem{Trev1}
  D.~Tong,
 {\it ``TASI lectures on solitons,''}
arXiv:hep-th/0509216.
  %%CITATION = HEP-TH/0509216;%%
%\cite{Tong:2008qd}

\bibitem{Trev2}
  D.~Tong,
  %``Quantum Vortex Strings: A Review,''
  arXiv:0809.5060 [hep-th].
  %%CITATION = ARXIV:0809.5060;%%




\bibitem{EINOKrev}
  M.~Eto, Y.~Isozumi, M.~Nitta, K.~Ohashi and N.~Sakai,
  %``Solitons in the Higgs phase: The moduli matrix approach,''
  J.\ Phys.\ A  {\bf 39} (2006) R315
  [arXiv:hep-th/0602170].
  %%CITATION = JPAGB,A39,R315;%%

\bibitem{mamayung}
  V.~Markov, A.~Marshakov and A.~Yung,
  %``Non-Abelian vortices in N = 1* gauge theory,''
  Nucl.\ Phys.\  B {\bf 709} (2005) 267
  [arXiv:hep-th/0408235].


\bibitem{misc}
  H.~J.~de Vega and F.~A.~Schaposnik,
  %``ELECTRICALLY CHARGED VORTICES IN NONABELIAN GAUGE THEORIES WITH
  %CHERN-SIMONS TERM,''
  Phys.\ Rev.\ Lett.\  {\bf 56} (1986) 2564;
  %%CITATION = PRLTA,56,2564;%%
  H.~J.~de Vega and F.~A.~Schaposnik,
  %``Vortices and electrically charged vortices in nonAbelian gauge theories,''
  Phys.\ Rev.\  D {\bf 34} (1986) 3206;
  %%CITATION = PHRVA,D34,3206;%%
  J.~Heo and T.~Vachaspati,
  %``Z(3) strings and their interactions,''
  Phys.\ Rev.\  D {\bf 58} (1998) 065011
  [arXiv:hep-ph/9801455];
  %%CITATION = PHRVA,D58,065011;%%
  F.~A.~Schaposnik and P.~Suranyi,
  %``New vortex solution in SU(3) gauge-Higgs theory,''
  Phys.\ Rev.\  D {\bf 62} (2000) 125002
  [arXiv:hep-th/0005109];
  %%CITATION = PHRVA,D62,125002;%%
  K.~Konishi and L.~Spanu,
  %``Non-Abelian vortex and confinement,''
  Int.\ J.\ Mod.\ Phys.\  A {\bf 18} (2003) 249
  [arXiv:hep-th/0106175].
  %%CITATION = IMPAE,A18,249;%%

\bibitem{kneipp}
  M.~A.~C.~Kneipp,
  %``BPS Z_N String Tensions, Sine and Casimir Laws and Integrable Field
  %Theories,''
  Phys.\ Rev.\  D {\bf 76} (2007) 125010
  [arXiv:0707.3791 [hep-th]];
  %%CITATION = PHRVA,D76,125010;%%
 M.~A.~C.~Kneipp,
  %``colour superconductivity, Z(N) flux tubes and monopole confinement in
  %deformed N = 2* super Yang-Mills theories,''
  Phys.\ Rev.\  D {\bf 69} (2004) 045007
  [arXiv:hep-th/0308086].
  %%CITATION = PHRVA,D69,045007;%%


\bibitem{witten}
  E.~Witten,
  %``Instantons, The Quark Model, And The 1/N Expansion,''
  Nucl.\ Phys.\  B {\bf 149}, 285 (1979).
  %%CITATION = NUPHA,B149,285;%%

\bibitem{zam1}
A.~B.~Zamolodchikov and A.~B.~Zamolodchikov,
  %``Factorized S-matrices in two dimensions as the exact solutions of  certain
  %relativistic quantum field models,''
  Annals Phys.\  {\bf 120}, 253 (1979).
  %%CITATION = APNYA,120,253;%%


\bibitem{zam2}
A.~B.~Zamolodchikov and A.~B.~Zamolodchikov,
  %``Massless Factorized Scattering And Sigma Models With Topological Terms,''
  Nucl.\ Phys.\  B {\bf 379}, 602 (1992).
  %%CITATION = NUPHA,B379,602;%%


\bibitem{zam3}
  V.~A.~Fateev, E.~Onofri and A.~B.~Zamolodchikov,
  %``The Sausage model (integrable deformations of O(3) sigma model),''
  Nucl.\ Phys.\  B {\bf 406} (1993) 521.
  %%CITATION = NUPHA,B406,521;%%

\bibitem{biaggio}
  B.~Lucini and M.~Teper,
  %``Confining strings in SU(N) gauge theories,''
  Phys.\ Rev.\  D {\bf 64}, 105019 (2001)
  [arXiv:hep-lat/0107007].
  %%CITATION = PHRVA,D64,105019;%%

  B.~Lucini, M.~Teper and U.~Wenger,
  %``Glueballs and k-strings in SU(N) gauge theories: Calculations with
  %improved operators,''
  JHEP {\bf 0406}, 012 (2004)
  [arXiv:hep-lat/0404008].
  %%CITATION = JHEPA,0406,012;%%

\bibitem{hk}
  C.~P.~Herzog and I.~R.~Klebanov,
  %``On string tensions in supersymmetric SU(M) gauge theory,''
  Phys.\ Lett.\  B {\bf 526} (2002) 388
  [arXiv:hep-th/0111078].
  %%CITATION = PHLTA,B526,388;%%

\bibitem{ds}
  M.~R.~Douglas and S.~H.~Shenker,
  %``Dynamics of SU(N) supersymmetric gauge theory,''
  Nucl.\ Phys.\  B {\bf 447} (1995) 271
  [arXiv:hep-th/9503163].
  %%CITATION = NUPHA,B447,271;%%

\bibitem{hsz}
  A.~Hanany, M.~J.~Strassler and A.~Zaffaroni,
  %``Confinement and strings in M{QCD},''
  Nucl.\ Phys.\  B {\bf 513} (1998) 87
  [arXiv:hep-th/9707244].
  %%CITATION = NUPHA,B513,87;%%

%\cite{Myers:1999ps}
\bibitem{myers}
  R.~C.~Myers,
  %``Dielectric-branes,''
  JHEP {\bf 9912} (1999) 022
  [arXiv:hep-th/9910053].
  %%CITATION = JHEPA,9912,022;%%


\bibitem{gsy}
 A.~Gorsky, M.~Shifman and A.~Yung,
  %``Non-Abelian Meissner effect in Yang-Mills theories at weak coupling,''
  Phys.\ Rev.\  D {\bf 71} (2005) 045010
  [arXiv:hep-th/0412082].
  %%CITATION = PHRVA,D71,045010;%%



\bibitem{dorey}
 N.~Dorey,
  %``S-duality, deconstruction and confinement for a marginal deformation of  N
  %= 4 SUSY Yang-Mills,''
  JHEP {\bf 0408} (2004) 043
  [arXiv:hep-th/0310117];
  %%CITATION = JHEPA,0408,043;%%
  N.~Dorey,
  %``A new deconstruction of little string theory,''
  JHEP {\bf 0407} (2004) 016
  [arXiv:hep-th/0406104].
  %%CITATION = JHEPA,0407,016;%%

\bibitem{da}
R.~P.~Andrews and N.~Dorey,
  %``Spherical deconstruction,''
  Phys.\ Lett.\  B {\bf 631} (2005) 74
  [arXiv:hep-th/0505107];
  %%CITATION = PHLTA,B631,74;%%
  R.~P.~Andrews and N.~Dorey,
  %``Deconstruction of the Maldacena-Nunez compactification,''
  Nucl.\ Phys.\  B {\bf 751} (2006) 304
  [arXiv:hep-th/0601098].
  %%CITATION = NUPHA,B751,304;%%

\bibitem{madore}
  J.~Madore,
  %``The fuzzy sphere,''
  Class.\ Quant.\ Grav.\  {\bf 9} (1992) 69.
  %%CITATION = CQGRD,9,69;%%


\bibitem{heterotic}
  M.~Edalati and D.~Tong,
  %``Heterotic vortex strings,''
  JHEP {\bf 0705} (2007) 005
  [arXiv:hep-th/0703045];
  %%CITATION = JHEPA,0705,005;%%
  D.~Tong,
  %``The quantum dynamics of heterotic vortex strings,''
  JHEP {\bf 0709} (2007) 022
  [arXiv:hep-th/0703235];
  %%CITATION = JHEPA,0709,022;%%
  M.~Shifman and A.~Yung,
  %``Heterotic Flux Tubes in N=2 SQCD with N=1 Preserving Deformations,''
  Phys.\ Rev.\  D {\bf 77} (2008) 125016
  [arXiv:0803.0158 [hep-th]];
  %%CITATION = PHRVA,D77,125016;%%
  M.~Shifman and A.~Yung,
  %``Large-N Solution of the Heterotic N=(0,2) Two-Dimensional CP(N-1) Model,''
  Phys.\ Rev.\  D {\bf 77} (2008) 125017
  [arXiv:0803.0698 [hep-th]].
  %%CITATION = PHRVA,D77,125017;%%

\bibitem{affleckhaldane}
  I.~Affleck,
  %``Exact Critical Exponents For Quantum Spin Chains, Nonlinear Sigma Models At
  %Theta = Pi And The Quantum Hall Effect,''
  Nucl.\ Phys.\  B {\bf 265} (1986) 409;
  %%CITATION = NUPHA,B265,409;%%
  I.~Affleck and F.~D.~M.~Haldane,
  %``CRITICAL THEORY OF QUANTUM SPIN CHAINS,''
  Phys.\ Rev.\  B {\bf 36} (1987) 5291;
  %%CITATION = PHRVA,B36,5291;%%
  I.~Affleck,
  %``Nonlinear sigma model at Theta = pi: Euclidean lattice formulation and
  %solid-on-solid models,''
  Phys.\ Rev.\ Lett.\  {\bf 66} (1991) 2429.
  %%CITATION = PRLTA,66,2429;%%



\bibitem{controzzi}
  D.~Controzzi and G.~Mussardo,
  %``On the mass spectrum of the two-dimensional O(3) sigma model with theta
  %term,''
  Phys.\ Rev.\ Lett.\  {\bf 92} (2004) 021601
  [arXiv:hep-th/0307143].
  %%CITATION = PRLTA,92,021601;%%

\bibitem{shankarread}
  R.~Shankar and N.~Read,
  %``THE THETA = pi NONLINEAR SIGMA MODEL IS MASSLESS,''
  Nucl.\ Phys.\  B {\bf 336} (1990) 457.
  %%CITATION = NUPHA,B336,457;%%


\bibitem{Nekrasov}
  N.~Nekrasov and A.~S.~Schwarz,
  %``Instantons on noncommutative R**4 and (2,0) superconformal six  dimensional
  %theory,''
  Commun.\ Math.\ Phys.\  {\bf 198} (1998) 689
  [arXiv:hep-th/9802068].
  %%CITATION = CMPHA,198,689;%%




\bibitem{AOSJ}
  M.~Alishahiha, Y.~Oz and M.~M.~Sheikh-Jabbari,
  %``Supergravity and large N noncommutative field theories,''
  JHEP {\bf 9911} (1999) 007
  [arXiv:hep-th/9909215].
  %%CITATION = JHEPA,9911,007;%%

\bibitem{BMM}
  J.~C.~Breckenridge, G.~Michaud and R.~C.~Myers,
  %``More D-brane bound states,''
  Phys.\ Rev.\  D {\bf 55} (1997) 6438
  [arXiv:hep-th/9611174].
  %%CITATION = PHRVA,D55,6438;%%

\bibitem{Btricky}
  C.~Bachas, M.~R.~Douglas and C.~Schweigert,
  %``Flux stabilization of D-branes,''
  JHEP {\bf 0005} (2000) 048
  [arXiv:hep-th/0003037];
  %%CITATION = JHEPA,0005,048;%%
  W.~Taylor,
  %``D2-branes in B fields,''
  JHEP {\bf 0007} (2000) 039
  [arXiv:hep-th/0004141];
  %%CITATION = JHEPA,0007,039;%%
  A.~Alekseev, A.~Mironov and A.~Morozov,
  %``On B-independence of RR charges,''
  Phys.\ Lett.\  B {\bf 532} (2002) 350
  [arXiv:hep-th/0005244];
  %%CITATION = PHLTA,B532,350;%%
  J.~G.~Zhou,
  %``D-branes in B fields,''
  Nucl.\ Phys.\  B {\bf 607} (2001) 237
  [arXiv:hep-th/0102178].
  %%CITATION = NUPHA,B607,237;%%




%\cite{Drukker:2005kx}
\bibitem{dfiol}
  N.~Drukker and B.~Fiol,
  %``All-genus calculation of Wilson loops using D-branes,''
  JHEP {\bf 0502}, 010 (2005)
  [arXiv:hep-th/0501109].
  %%CITATION = JHEPA,0502,010;%%


%\cite{Yamaguchi:2006tq}
\bibitem{yama}
  S.~Yamaguchi,
  %``Wilson loops of anti-symmetric representation and D5-branes,''
  JHEP {\bf 0605}, 037 (2006)
  [arXiv:hep-th/0603208].
  %%CITATION = JHEPA,0605,037;%%

\bibitem{malpol}
  S.~A.~Hartnoll and S.~Prem Kumar,
  %``Multiply wound Polyakov loops at strong coupling,''
  Phys.\ Rev.\  D {\bf 74}, 026001 (2006)
  [arXiv:hep-th/0603190].
  %%CITATION = PHRVA,D74,026001;%%

%\cite{Callan:1997kz}
\bibitem{bion}
  C.~G.~Callan and J.~M.~Maldacena,
  %``Brane dynamics from the Born-Infeld action,''
  Nucl.\ Phys.\  B {\bf 513}, 198 (1998)
  [arXiv:hep-th/9708147].
  %%CITATION = NUPHA,B513,198;%%

%\cite{Witten:1998xy}
\bibitem{wittenbaryon}
  E.~Witten,
  %``Baryons and branes in anti de Sitter space,''
  JHEP {\bf 9807}, 006 (1998)
  [arXiv:hep-th/9805112].
  %%CITATION = JHEPA,9807,006;%%

%\cite{Callan:1998iq}
\bibitem{calguij1}
  C.~G.~.~Callan, A.~Guijosa and K.~G.~Savvidy,
  %``Baryons and string creation from the fivebrane worldvolume action,''
  Nucl.\ Phys.\  B {\bf 547}, 127 (1999)
  [arXiv:hep-th/9810092].
  %%CITATION = NUPHA,B547,127;%%

%\cite{calguij2}
\bibitem{calguij2}
  C.~G.~.~Callan, A.~Guijosa, K.~G.~Savvidy and O.~Tafjord,
  %``Baryons and flux tubes in confining gauge theories from brane actions,''
  Nucl.\ Phys.\  B {\bf 555}, 183 (1999)
  [arXiv:hep-th/9902197].
  %%CITATION = NUPHA,B555,183;%%

%\cite{Hartnoll:2004yr}
\bibitem{sean}
  S.~A.~Hartnoll and R.~Portugues,
  %``Deforming baryons into confining strings,''
  Phys.\ Rev.\  D {\bf 70}, 066007 (2004)
  [arXiv:hep-th/0405214].
  %%CITATION = PHRVA,D70,066007;%%

%\cite{Hanany:1996ie}
\bibitem{hwitten}
  A.~Hanany and E.~Witten,
  %``Type IIB superstrings, BPS monopoles, and three-dimensional gauge
  %dynamics,''
  Nucl.\ Phys.\  B {\bf 492}, 152 (1997)
  [arXiv:hep-th/9611230].
  %%CITATION = NUPHA,B492,152;%%

\bibitem{inprog}

R. Auzzi and S. P. Kumar, Work in progress.









\end{thebibliography}
\end{document}